**Dynamics of monohydroxy alcohols with chain-like structures: Hydrogen bonding lifetime, chain swapping, and Debye process**

Shiwang Cheng* and Shalin Patil

Department of Chemical Engineering and Materials Science, Michigan State University, East Lansing, MI 48824, USA

## Abstract

By assuming reversible H-bonding association and dissociation, this work provides a detailed description of the supramolecular structure and dynamics of monohydroxy alcohols (MAs) within the framework of a recently proposed living chain model (LCM). Structurally, reversible H-bonding leads to a single exponential distribution of the molar concentration of the supramolecular chain with length $N$, $c(N) \sim \exp(-N/\bar{N})$, with $\bar{N}$ being the characteristic chain length following $\bar{N} \sim c_0^{1/2}$ and $c_0$ being the concentration of MA participating into the supramolecular chains. Dynamically, reversible H-bonding enables supramolecular chain breakage and recombination, which modifies the relaxation time of the supramolecular chains. In addition to the structural relaxation, $\tau_\alpha$, and the Debye relaxation, $\tau_D$, two other relaxation times are revealed: the characteristic time of the chain breakage, $\tau_B$, and the lifetime of a H-bonding, $\tau_H$. The interplay among these four-time scales, $\tau_\alpha$, $\tau_B$, $\tau_D$, and $\tau_H$, defines *five* distinct dynamics regimes. In Regimes I and V, $\tau_\alpha > \tau_H$ holds, and no supramolecular chains form. In Regimes II and IV ($\tau_H \approx \tau_D > \tau_B \approx \tau_\alpha$), supramolecular chains form and give a Debye relaxation. The characteristic chain length scales as $\bar{N} \approx \tau_D/\tau_\alpha$. In these two regimes, the H-bonding lifetime controls the Debye process. In Regime III, $\tau_H > \tau_D \gg \tau_B > \tau_\alpha$ and large supramolecular chains form with $\bar{N} \approx (\tau_D^2/(\tau_B\tau_\alpha))^{1/2}$. In all regimes with supramolecular chain formation, the Debye relaxation comes from the overall chain end-to-end dipole reorientation and scales with $\bar{N}$. Excellent agreements between experiments and the LCM have been observed, leading to quantitative descriptions of the dielectric and linear viscoelastic properties of MAs. These results thus establish a theoretical framework linking reversible pairwise H-bonding interactions to supramolecular structures, dynamics, and macroscopic properties of MAs.

* Corresponding Author. Email Address: chengsh9@msu.edu

## 1. Introduction

The influence of hydrogen bonding (H-bonding) on the structures and dynamics of monohydroxy alcohols (MAs) has been a topic of active study for more than a century.[1-25] In particular, MAs give a strong dielectric absorption, i.e. the Debye process, at times longer than their structural relaxation for glass transition.[6, 22, 26] For a long time, the single-relaxation mode Debye process has been observed only in dielectric measurements.[6] Recent studies reveal clear correspondences of the Debye relaxation of MAs the depolarized dynamic light scattering (DDLS)[9, 27, 28] and linear rheology[7, 10, 16, 17, 20, 29]. Various theoretical models and understandings have been proposed for the origin of the Debye relaxation, including the H-bonding lifetime,[14, 30, 31] a sequential switch of H-bonding,[32] the interfacial polarization between the hydrophilic H-bonding clusters and the adjacent hydrophobic alkyl groups,[33] a molecular rearrangement of the H-bonding within the clusters of hydroxyl groups,[2] the chain-$g_k$ fluctuation with $g_k$ being the Kirkwood-Fröhlich factor,[4] the transient chain model (TCM),[1, 3, 7, 34] the density fluctuation,[20] the dipole-dipole cross correlation,[9, 11, 13, 15, 19] the migration of defects in the supramolecular structure,[35] and the living chain model (LCM)[16, 17]. Despite the significant progress on the characterizations of Debye process in MAs, a key question remains: what is the relationship between the reversible H-bonding interaction and the Debye relaxation of MAs?

In addition, Debye relaxation has been widely observed in water[36-38], finds its signature in supramolecular polymeric liquids[39], and recently has been observed in non-hydrogen bonding polar molecular liquids[19, 40, 41]. Therefore, a clear elucidation of the nature of the H-bonding association of MAs and their relationship with Debye relaxation could help understand the structures, dynamics, and dielectric properties of other types of associative liquids.[42-46]

Due to the single-relaxation mode of the Debye process, early studies attribute the H-bonding lifetime time as the origin of the Debye process.[14, 30] However, this does not explain the super-Arrhenius temperature dependence of the Debye relaxation time, $\tau_D$, in the supercooled region, and the absence of a strong Debye relaxation in polyalcohols.[47-52] Recently, Gainaru et al conducted new nuclear magnetic resonance (NMR) spectroscopy measurements,[1, 3, 34, 53] and revealed a time scale associated with the H-bonding dynamics at times between the structural relaxation time, $\tau_\alpha$, and the Debye time, $\tau_D$. Based on the new observations, Gainaru et al proposed a transient chain model (TCM)[1] that emphasizes a supramolecular chain formation in MAs. The authors argued that the Debye relaxation is from the supramolecular chain end-to-end vector reorientation, which is achieved through the H-bonding association and dissociation at chain ends. Other mechanisms, including the chain-$g_k$ fluctuation mechanism[4] and the density fluctuation mechanism[20], also suggest the supramolecular chain end-to-end vector reorientation as the leading mechanism of the Debye relaxation.

In the meanwhile, theoretical calculations suggest that the dipole-dipole cross-correlation mechanism (DDCM) can give rise to a Debye process of polar liquids at times longer than the structural relaxation at $g_k > 1$.[11] One fundamental difference between the DDCM and TCM is that DDCM does not involve supramolecular chain formation for Debye relaxation. Due to the experimental challenges to quantify the dipole-dipole cross-correlation, an explicit relationship between the DDCM and the Debye relaxation remains a topic of on-going discussion.[13, 15, 18, 19, 40, 54]

Very recently, Cheng et al have conducted novel rheo-dielectric measurements to characterize the dynamics of MAs, including the Debye relaxation.[16] A strong influence of the external shear on the dynamics of MAs has been observed. The external shear speeds up both $\tau_\alpha$ and $\tau_D$. The shear-induced shifts in $\tau_\alpha$ are noticeably larger than $\tau_D$, leading to a shear-induced further separation between $\tau_D$ and $\tau_\alpha$. Furthermore, the shear-induced shifts in $\tau_\alpha$ and $\tau_D$ depend on the applied shear rate $\dot{\gamma}$. Quantitative analyses revealed a scaling behavior $\frac{\tau_D(\dot{\gamma})}{\tau_D(0)} \sim \left(\frac{\tau_\alpha(\dot{\gamma})}{\tau_\alpha(0)}\right)^{1/2}$ and a universal Arrhenius temperature dependence of $\tau_D^2(T)/\tau_\alpha(T) \sim \exp(\Delta E_a/(RT))$ that hold for various MAs with different H-bond types and alkyl structures, where $\Delta E_a$ is the apparent activation energy of $\tau_D^2(T)/\tau_\alpha(T)$.[16] In addition, new stress relaxation measurements and the relaxation time distribution analyses have revealed rich dynamics of MAs at intermediate times between $\tau_\alpha$ and $\tau_D$, including the normal modes of the supramolecular chains and the emergence of new dynamics modes at the intermediate times.[17] According to the recent study, these processes at the intermediate times come from the chain breakage process with a characteristic time $\tau_B$, which matches well with the H-bonding exchange times revealed by the NMR spectroscopy.[17]

These new observations motivate the development of a living chain model (LCM).[16, 17] The LCM emphasizes the reversibility of the H-bonding association and dissociation and the effective breakage of the supramolecular chains. In the limit of fast chain breakage, $\tau_B \ll \tau_D$, active chain breakage and recombination speed up the end-to-end vector reorientation. Although LCM offers an understanding of the dynamics of the transient supramolecular chains with initial success, a detailed comparison between LCM and experiments has yet to be conducted. Furthermore, previous discussions of LCM focus primarily on the cases where the intermediate time, $\tau_B$, separates largely from $\tau_\alpha$,[16,17] leaving other dynamics regimes largely unexplored.

In this work, we present a detailed derivation of the LCM for the structures and dynamics of molecular liquids with pair-wise reversible interactions and discuss its application for MAs. We compare the predictions of LCM with the dielectric and linear viscoelastic properties of MAs. Excellent agreements between LCM model and experiments are observed, from which one obtains key parameters to describe the dynamics of MAs, including the average chain sizes, $\bar{N}$, the H-bonding lifetime, $\tau_H$, and the relationship between $\tau_H$, $\tau_D$, $\tau_B$, and $\tau_\alpha$. Due to the different temperatures dependence of $\tau_H$, $\tau_D$, $\tau_B$, and $\tau_\alpha$, one finds two temperatures, $T_I$ and $T_{IV}$ ($T_I > T_{IV}$) where $\tau_H = \tau_\alpha$ holds. Similarly, there are two temperatures, $T_{II}$ and $T_{III}$ ($T_{II} > T_{III}$) where $\tau_B = \tau_\alpha$ holds. Depending on the relative values of $\tau_H$, $\tau_D$, $\tau_B$, and $\tau_\alpha$, five distinct dynamics regimes can be categorized: **(i)** Regime I at $T \geq T_I$. In this regime, H-bonding lifetime is shorter than structural relaxation time. The dipolar reorientation of individual MA controls the structural relaxation, and no Debye process can be observed. **(ii)** Regime II at $T_{II} \leq T < T_I$. In this regime, supramolecular structures form with a dynamics size of $N_R = \sqrt{\tau_B/\tau_\alpha} = 1$. Structural relaxation and Debye relaxation are two separated processes, and chain swapping takes place at the time scale of structural relaxation. The supramolecular chain sizes follow $\bar{N} \approx \tau_D/\tau_\alpha$, and $\tau_D \approx \tau_H$ holds, i.e., the H-bonding lifetime controls the Debye process in this regime. **(iii)** Regime III at $T_{III} \leq T < T_{II}$. Large supramolecular structures form and $\tau_H > \tau_D > \tau_B > \tau_\alpha$. $N_R = \sqrt{\tau_B/\tau_\alpha} > 1$ and the characteristic chain length follows $\bar{N} \approx (\tau_D^2/(\tau_B\tau_\alpha))^{1/2}$. The Debye time $\tau_D \approx \tau_H/N_R < \tau_H$, i.e. the Debye time is shorter than the H-bonding lifetime. At the same time, $\tau_D \approx \tau_\alpha N_R \bar{N} < \tau_\alpha \bar{N}^2$, suggesting the active H-bonding association and dissociation speed up the end-to-end reorientation of supramolecular chains. **(iv)** Regime IV at $T_{IV} \leq T < T_{III}$. This can be viewed as a low temperature analog of Regime II, where a strong rise of the configurational entropic barrier for structural relaxation plays an essential role. **(v)** Regime V at $T < T_{IV}$ which can be viewed as the low-temperature analog of Regime I, and molecular cooperativity controls structural relaxation. Thus, Debye process appears in Regimes II, III, and IV, and is associated with the end-to-end vector reorientation of supramolecular chains. Furthermore, LCM leads to a quantitative description of the dielectric relaxation amplitude of the Debye relaxation, $\Delta\varepsilon_D$, and $g_k$, as well as their relationship with $\tau_H$ and $\bar{N}$. These results offer new insights into the multiscale supramolecular dynamics of MAs and elucidate the role of H-bonding interaction in governing their dielectric and viscoelastic properties.

## 2. The living chain model (LCM)

### 2.1 Supramolecular structure formation of MAs.

The LCM assumes the reversibility of the H-bonding interaction at *any* H-bonding site of the supramolecular structures. **Figure 1** presents the fundamental chemical reaction pathways of the H-bond formation and destruction, where $k_1$ and $k_2$ are the corresponding reaction rate constants, and $m$ and $n$ are the numbers of repeating units. The equilibrium reaction constant of the H-bonding association and dissociation is $K_{eq} = k_1/k_2$. Depending on the relative values of $k_1$ and $k_2$, the chemical reaction can either favor the supramolecular structure formation or the supramolecular fragmentation.

Assuming the molar concentration of supramolecular chain with length $N$ as $c(N)$, one can write down the materials balance equation:[55]

$$\frac{dc(N)}{dt} = -k_2 N c(N) - k_1 c(N) \int_0^{\infty} c(N')dN' + 2k_2 \int_N^{\infty} c(N')dN'$$

$$+ k_1 \int_0^{\infty} \int_0^{\infty} c(N')c(N'')\delta(N' + N'' - N)dN'dN'' \qquad (1)$$

The first term on the right represents the contribution of fragmentation of length $N$, the second term represents the chain $N$ lengthening, the third time is the generation of supramolecular structures with length $N$ through fragmentation of larger sizes, and the fourth term describes the supramolecular structures combination to form length $N$, where $\delta(x)$ is the Kronecker delta function. In the steady-state, $\frac{dc(N)}{dt} = 0$ holds and a solution to the steady-state equation is:

$$c(N) = \frac{2k_2}{k_1} \exp\left(-\frac{N}{\overline{N}}\right) \qquad (2)$$

where

$$\overline{N} = \sqrt{\frac{c_0 k_1}{2k_2}} = \sqrt{\frac{c_0 K_{eq}}{2}} \qquad (3)$$

and $c_0$ is the total molar concentration of monomers in the supramolecular chain structures. **Eqn 3** suggests that the characteristic supramolecular sizes of MAs and their sizes distribution depend on $K_{eq}$. Similar scaling of $\overline{N} \sim c_0^{1/2}$ has been obtained for equilibrium polymerization with free activation.[56] The volume fraction of supramolecular structures with size $N$ is

$$\varphi(N) = \epsilon\phi(N) = \frac{\epsilon N c(N)}{\sum_{i=1}^{\infty} i c(i)} \qquad (4)$$

where $\epsilon$ is the volume fraction of chains (including $N = 1$) and $\phi(N) = \frac{Nc(N)}{\sum_{i=1}^{\infty} ic(i)}$ is the volume fraction of the chain length $N$ within the supramolecular chain population.

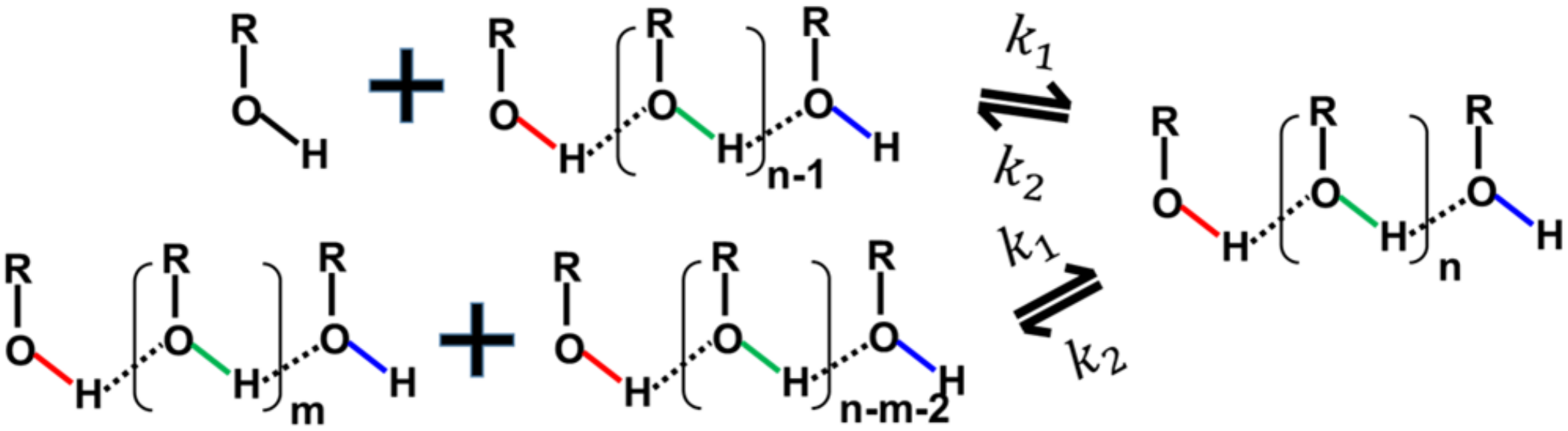

**Figure 1.** A schematic illustration of the hydrogen bonding (H-bonding) association and dissociation of monohydroxy alcohol (MA). The reaction rate constants of H-bonding association and dissociation are $k_1$ and $k_2$, respectively. ROH represents monohydroxy alcohol with R being the alkyl group. $n$ and $m$ are the numbers of repeating units.

### 2.2 Dynamics of the supramolecular chains.

Due to the supramolecular formation, dynamics of MAs are far richer than the non-hydrogen bonding molecular liquids. The H-bonding lifetime is $\tau_H = \frac{1}{k_2}$, and the structural relaxation, $\tau_\alpha$, of MAs mimics segmental dynamics of polymers. The glass transition temperature $T_g$ and the zero-shear viscosity $\eta$ of MAs are much higher than the corresponding alkanes.[57, 58] As discussed in a following section, these supramolecular structures exhibit intriguing dynamics slower than $\tau_\alpha$, such as the chain breakage and recombination, $\tau_B$, the terminal relaxation from linear rheology, $\tau_f$, and the Debye relaxation from dielectric spectroscopy, $\tau_D$.[6] The flow time or the Debye time quantifies the end-to-end reorientation of the supramolecular chains, and $\tau_f \approx \tau_D$ holds.[4, 7, 10, 16, 17, 20]

Assuming each H-bond has an equal probability to dissociate, a supramolecular chain with $N$ molecules, i.e. $(N-1)$ H-bonds, breaks up at $\tau_B = \frac{\tau_H}{N-1} = \frac{1}{k_2(N-1)} \sim \frac{1}{k_2 N}$ at $N >> 1$. The chain breakage time is a factor of $(N-1)$ smaller than the H-bond lifetime as a dissociation of any of the $(N-1)$ bonds leads to a chain breakage. Note that the concept of reversibility and effective breakup of associative bonds has been actively explored in dynamics of associative polymers[59-61] and wormlike micelles[62-64]. Unlike wormlike micelles, no signs of entanglement have been observed in the linear viscoelastic properties of MAs.[7, 10] Therefore, we do not consider entanglement effect in the discussion.

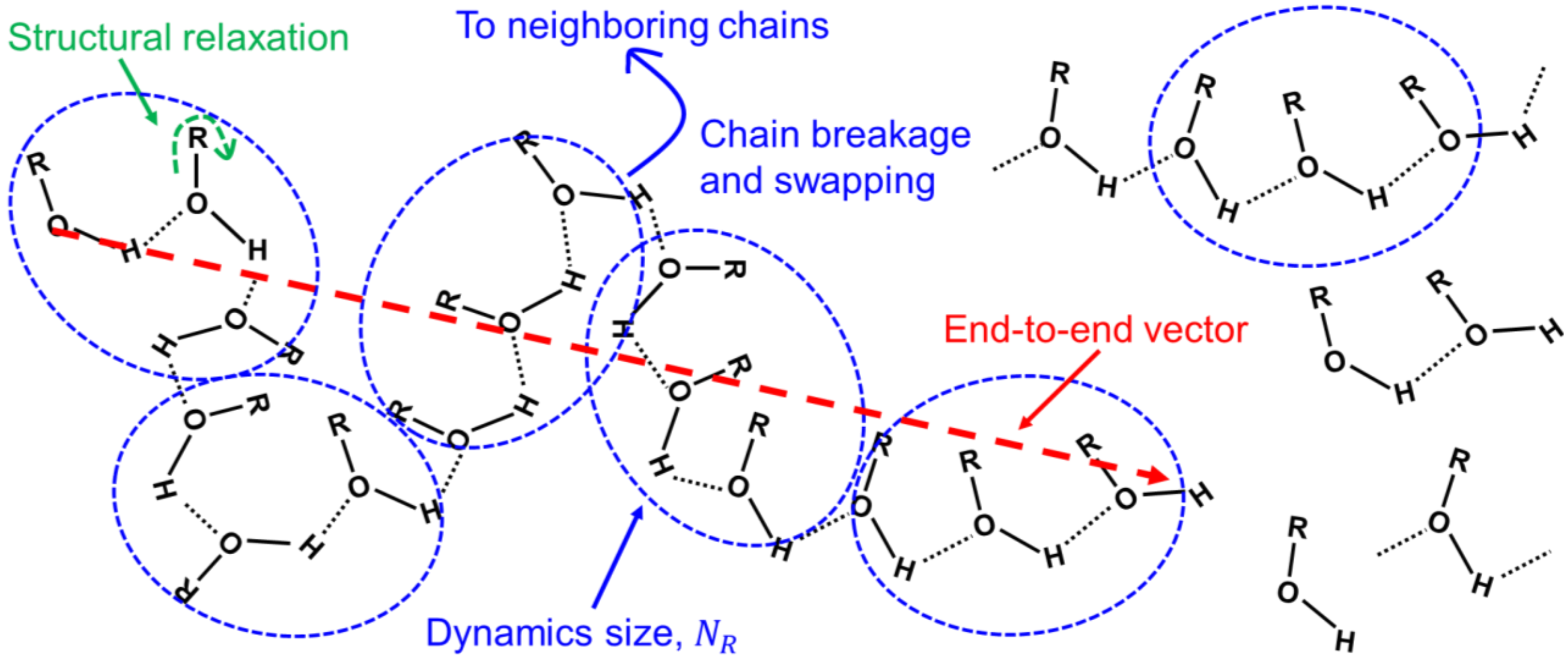

**Figure 2** An illustration of the multiscale dynamics of monohydroxy alcohols (MAs, ROH) with chain structures. The dotted black lines among MA represents the hydrogen bonding (H-bonding). There is a distribution of chain length, $N$, from monomer ($N = 1$) to long chains. Reorientation of the repeat unit or the monomer provides structural relaxation (the green dashed arrow). The blue dashed ellipsoids represent the dynamics sizes with $N_R$ monomers in supramolecular structures, which serve as an elementary unit for chain swapping. Chains short than $N_R$ do not undergo chain swapping before their end-to-end vector reorientation. Chains longer than $N_R$ undergo the end-to-end vector (the red dashed vector) reorientation through multiple chain breakage and chain swapping. ROH represents monohydroxy alcohol with R being the alkyl group.

The supramolecular chains of different lengths, $N$, anticipate multiscale dynamics of MAs (**Figure 2**). At times longer than the structural relaxation and shorter than the chain breakage, i.e. $\tau_\alpha < t \leq \tau_B$, no chain breakage takes place and the supramolecular chains exhibit Rouse dynamics.[65] For short chains with length $N$, if $\tau_B \geq \tau_\alpha N^2$, the dynamics of these short chains behave like covalently bonded polymers. This gives a critical chain size at $\tau_B(N_c) \approx \frac{\tau_H}{N_c} \approx \tau_\alpha N_c^2$ and $N_c \approx (\tau_H/\tau_\alpha)^{1/3}$. At $N \leq N_c$, the entire supramolecular chains (i.e. oligomer) follow Rouse dynamics, and no chain breakage takes place before their terminal time. At $N > N_c$, the longest Rouse mode that can survive before a chain breakage event is the one with $N_R$ monomers, where $N_R \approx (\tau_B/\tau_\alpha)^{\frac{1}{2}} = (N_c/N)^{\frac{1}{2}} N_c < N_c$. Beyond the chain breakage time, $\tau_B$, chain breakage and active chain swapping must take place.[17]

Conceptually, each chain swap can lead to a block of $N_R$ monomers to dissipate their orientational order. Hence, $N_R$ serves as an elementary unit, i.e., the dynamics size, for the dynamics of the supramolecular chain. A chain with length $N$ needs to undergo, on average, $N/N_R$ time chain swapping to achieve its overall end-to-end reorientation. This gives a terminal time $\tau_f = (N/N_R) \times \tau_B = \tau_\alpha N_R N < \tau_\alpha N^2 = \tau_R$, where $\tau_R$ is the Rouse time of the entire supramolecular chain in the absence of chain breakage. Thus, the chain breakage and chain swapping facilitate the end-to-end vector reorientation. Since $\tau_f = (N/N_R) \times \tau_B \approx \frac{1}{k_2 N_R}$, the terminal time can also be viewed as the breakage time of the dynamics size, which is an argument Cates provided in the living polymer model for entangled polymers.[55] Furthermore, $\tau_f = \tau_\alpha N_R N = (\tau_B \tau_R)^{1/2}$, which is the geometrical average of the chain breakage time and the Rouse time of the supramolecular chain.[16] Similar scaling on the terminal relaxation time of unentangled living polymers have been discussed in Cates et al's recent study.[64]

The above analyses suggest a critical chain size, $N_c \approx (\tau_H/\tau_\alpha)^{1/3}$. At $N \leq N_c$, the dynamics of supramolecular chains follow Rouse dynamics. At $N > N_c$, chain breakage takes place at the intermediate time $\tau_B$, along with a dynamics size, $N_R \approx (\tau_B/\tau_\alpha)^{1/2}$. The cases of $N \leq N_c$ can be directly addressed through an analogy to the normal mode analysis of type-A polymer following Rouse dynamics.[66, 67] At $N > N_c$, one needs to consider the contributions of both the short chain and the long chains.

At $\bar{N} > N_c$, supramolecular chains behave like covalently bonded polymers at $\tau_\alpha < t \leq \tau_B$ as no chain breakage takes place. The H-bonding induced supramolecular chains have an *accumulated dipole* along the backbone, which can be captured by BDS, in analogy to a type-A polymer.[67] Thus, BDS spectra can capture the dielectric active normal modes of the supramolecular chain up to a Rouse bead $N_R \approx \sqrt{\tau_B/\tau_\alpha}$. At the same time, linear rheology could capture Rouse scaling, $G(t) \sim t^{-1/2}$ at $\tau_\alpha < t \leq \tau_B$. At $\tau_f \geq t > \tau_B$, supramolecular chains with sizes larger than $N_R$ monomers undergo active chain swapping. We would like to emphasize that the chain-swapping itself produces negligible variations in the net dipole moment because the total polarization of the system comes from a *vector* summation of individual dipoles. At the same time, the active chain breakage and recombination truncate the long-time Rouse mode of the supramolecular chains, leading to a stand out of the normal modes shorter than $\tau_B$.[17]

Since $N_R$ is the elementary dynamics size for chain swapping, one can divide the dynamics contributions to the intermediate frequencies as the following: (i) short chains at $N \leq N_R$ with no chain breakage, and (ii) long chains at $N > N_R$ with chain breakage and chain swapping. The relaxation processes between $\tau_\alpha$ and $\tau_D$ (or $\tau_f$) thus come from the following two sources: **(i)** the fastest $N_R$ normal modes of the long supramolecular chains with $N > N_R$; and **(ii)** the Rouse modes of short supramolecular structures with $N \leq N_R$. As shown in the discussion section, although the long supramolecular chain is a majority component within the supramolecular chains ($\phi(N > N_R){\sim}0.90 - 0.98$), the short chains of $N \leq N_R$ might contribute a significant portion of the dielectric relaxation amplitude at intermediate frequencies. Furthermore, the end-to-end dipole reorientation of the whole supramolecular chain accomplishes at $t \approx \tau_f \approx \tau_D$. It takes an average of $\bar{N}/N_R$ chain swapping to achieve the full reorientation of the end-to-end vector of a supramolecular chain with length $\bar{N}$. When $\bar{N}/N_R = \tau_f/\tau_B \gg 1$, $\tau_D$ follows a sharp Poisson distribution regardless of the distribution function of $\tau_B$.[68] Consequently, the terminal relaxation of the long-chain supramolecular structures follows a Debye relaxation in dielectric measurements[16] and a single-relaxation time Maxwell model in rheological measurement[62]. From this perspective, the Debye process in LCM is a *mathematical* outcome from the involvement of a large amount of chain swapping to achieve the overall end-to-end vector reorientation. In rheological measurements, each chain swapping leads to a drop in modulus of $\frac{\rho\varphi RT}{N_R M_0}$ with $\rho$ being the mass density of the MAs, $\varphi$ the volume fraction of the supramolecular chains, $R$ the gas constant, $T$ the absolute temperature, and $M_0$ the molecular weight of each alcohol molecule. Therefore, there is a drop in modulus by a factor of $\bar{N}/N_R$ from $\frac{\rho\varphi RT}{N_R M_0}$ to $\frac{\rho\varphi RT}{\bar{N} M_0}$ between $\tau_B$ and $\tau_f$. Since $\tau_f/\tau_B = \bar{N}/N_R$, this analysis gives a scaling of stress relaxation, i.e., $G(t){\sim}t^{-1}$ at $\tau_B < t < \tau_f$.

## 2.3 Modeling the dielectric and linear viscoelastic properties.

The complex dielectric permittivity, $\varepsilon^*(\omega)$, of MAs is composed of the end-to-end vector reorientation (the Debye process), $\varepsilon_D^*(\omega)$; the active normal modes of long chains with a dynamics size shorter than $N_R$, $\varepsilon_L^*(\omega)$; the dielectric active normal modes of short chains with $N \leq N_R$, $\varepsilon_S^*(\omega)$; and a glassy mode, $\varepsilon_\alpha^*(\omega)$:

$$\varepsilon^*(\omega) - \varepsilon_\infty = \varepsilon_D^*(\omega) + \varepsilon_L^*(\omega) + \varepsilon_S^*(\omega) + \varepsilon_\alpha^*(\omega) \qquad (5)$$

with $\varepsilon_\infty$ being the dielectric permittivity at the high-frequency limit and $\omega$ the angular frequency. We do not consider the DC conductivity and the electrode polarization process in the discussion since they show up at frequencies much lower than these molecular processes of MAs. The individual complex dielectric functions are:

$$\varepsilon_D^*(\omega) = \frac{\varphi(N > N_R)\Delta\varepsilon_{\bar{N}}}{1 + i\omega\tau_D} \qquad (6)$$

$$\varepsilon_L^*(\omega) = \frac{8\varphi(N > N_R)\Delta\varepsilon_{\bar{N}}}{\pi^2}\sum_{p\geq\bar{N}/N_R}^{\bar{N}}\left(\frac{1}{p^2}\frac{1}{1 + i\omega\tau_p}\right) \quad (p{:}\,odd) \quad (7)$$

$$\varepsilon_S^*(\omega) = \sum_{i=2}^{N_R}\left(\frac{8\varphi(i)\Delta\varepsilon_i}{\pi^2}\sum_{p=1}^{i}\left(\frac{1}{p^2}\frac{1}{1 + i\omega[\tau_\alpha(i/p)^2]}\right)\right) \quad (p{:}\,odd) \quad (8)$$

$$\varepsilon_\alpha^*(\omega) = \frac{\Delta\varepsilon_\alpha}{[1 + (i\omega\tau_\alpha)^{\beta_0}]^{\gamma_0}} \qquad (9)$$

Practically, we use

$$\varepsilon_L^*(\omega) \approx \frac{8\varphi(N > N_R)\Delta\varepsilon_{\bar{N}}}{\pi^2}\sum_{q=1}^{N_R}\left(\frac{q^2}{\bar{N}^2}\frac{1}{1 + i\omega\tau_\alpha q^2}\right) \qquad (10)$$

to replace **Eqn. 7**, which describes the dielectric active normal modes involve Rouse modes smaller than the dynamics size, $N_R$. In the above equations, $\Delta\varepsilon_{\bar{N}} = \frac{\rho N_A\langle\mu_\parallel^2\rangle F C_{\bar{N}} g_{\bar{N}}}{3\varepsilon_0 k_B T M_0}$ is the dielectric relaxation amplitude of the longest normal mode[69, 70] with $g_{\bar{N}}$ being the Kirkwood-Fröhlich factor of a chain $\bar{N}$ and $C_{\bar{N}}$ a constant connecting the monomer dipole moment to the chain end-to-end dipole, $\Delta\varepsilon_i = \frac{\rho N_A \langle\mu_{\parallel}^2\rangle F C_i g_i}{3\varepsilon_0 k_B T M_0}$ with $g_i$ being the Kirkwood-Fröhlich factor of chain length $i$ and $C_i$ a constant connecting the monomer dipole moment to the chain end-to-end dipole at chain length $i$, $\varphi(i)$ is the volume fraction of chains with length $i$, $\varphi(N > N_R)$ is the volume fraction of chains longer than $N_R$, $N_A$ is the Avogadro's number, $\mu_{\parallel} \approx 1.5\ D$ is the dipole moment of the hydroxyl group,[1] $\varepsilon_0$ is the vacuum permittivity, $k_B$ is the Boltzmann constant, $T$ is the absolute temperature, $M_0$ is the molecular weight of the MA, and $F = \frac{\varepsilon_S(\varepsilon_\infty+2)^2}{3(2\varepsilon_S+\varepsilon_\infty)}$ is the Onsager factor for the supramolecular process with $\varepsilon_s$ being the static dielectric constant. Given the reversibility of the H-bonding that gives the flexibility of the supramolecular chain, we assume $C_i \approx C_{\bar{N}}$ and $g_i \approx g_{\bar{N}}$ as a first approximation. In addition, $\tau_p = \tau_\alpha(\bar{N}/p)^2$ is the $p$th Rouse mode that involves a block of $\bar{N}/p$ monomer. Note that only the odd Rouse modes are dielectric active. $\Delta\varepsilon_\alpha = \frac{\rho N_A \langle\mu_{\perp}^2\rangle F g_\alpha}{3\varepsilon_0 k_B T M_0}$ is the dielectric relaxation strength of the structural relaxation, $\mu_{\perp} \approx 0.74\ D$ is the dipole moment of O-R bond,[1, 71] $g_\alpha$ is the Kirkwood-Fröhlich factor of the structural relaxation mode, and $\beta_0$ and $\gamma_0$ the shape parameters of the Havriliak-Negami (HN) function. Following a recent discovery of a generic shape of the structural relaxation of molecular liquids, we choose $\beta_0 = 0.8$ and $\gamma_0 = 0.6$ for the structural relaxation of MAs.[41]

At the same time, one can write down the complex modulus, $G^*(\omega)$:

$$G^*(\omega) = G_g^*(\omega) + G_L^*(\omega) + G_S^*(\omega) \qquad (11)$$

where $G_g^*(\omega)$ is the glassy mode, $G_L^*(\omega)$ represents the viscoelastic properties of chain longer than $N_R$, and $G_S^*(\omega)$ is the viscoelastic contribution of chains shorter than or equal to $N_R$. The corresponding expressions are:

$$G_g^*(\omega) = \frac{i\omega G_\infty \tau_\alpha}{1 + i\omega\tau_\alpha} \qquad (12)$$

$$G_S^*(\omega) = \sum_{j=2}^{N_R} \left( \frac{i\omega\rho RT\varphi(j)}{jM_0} \sum_{p=2}^{j} \frac{\tau_\alpha (j/p)^2}{1 + i\omega\tau_\alpha (j/p)^2} \right) \qquad (13)$$

$$G_L^*(\omega) = \frac{i\omega\rho RT\varphi(N > N_R)}{N_R M_0} \left( \sum_{p=2}^{N_R} \frac{\tau_B/p^2}{1 + i\omega\,\tau_B/p^2} + \int_{\tau_B}^{\infty} \frac{(\tau_B/\tau)^1}{i\omega + 1/\tau_f + 1/\tau} \frac{d\tau}{\tau} \right) \qquad (14)$$

Note that all the $p = 1$ mode of **Eqn. 13** goes to **Eqn. 12**, and we use a simple-mode Maxwell model (**Eqn. 12**) to represent the glassy mode contribution for the simplicity of the discussions, which captures well the low-frequency side of the viscoelastic spectra at $\omega < 1/\tau_\alpha$. **Eqn. 13** accounts for the Rouse contribution of the short chains ($N \leq N_R$), and **Eqn. 14** is an outcome of the Fourier transform of the stress relaxation function, $G(t)$, of MAs, where a detailed derivation of $G(t)$ has also been provided in a previous publication.[17] In particular, the first term in the bracket on the right side of **Eqn. 14** follows a Rouse description of the fastest $N_R - 1$ Rouse modes, and the second term describes the chain swapping dynamics.

At $\tau_B \approx \tau_\alpha$, $N_R \approx 1$. the complex dielectric function reduces to

$$\varepsilon^*(\omega) - \varepsilon_\infty = \varepsilon_D^*(\omega) + \varepsilon_\alpha^*(\omega) = \frac{\varphi(N \geq N_R)\Delta\varepsilon_{\bar{N}}}{1 + i\omega\tau_D} + \frac{\Delta\varepsilon_\alpha}{[1 + (i\omega\tau_\alpha)^{\beta_0}]^{\gamma_0}} \qquad (15)$$

At the same time, the complex moduli reduce to:

$$G^*(\omega) = \frac{i\omega G_\infty \tau_\alpha}{1 + i\omega\tau_\alpha} + \frac{i\omega\rho\varphi(N \geq N_R)RT}{M_0} \int_{\tau_\alpha}^{\infty} \frac{(\tau_\alpha/\tau)^1}{i\omega + 1/\tau_f + 1/\tau} \frac{d\tau}{\tau} \qquad (16)$$

**Eqns. 5-16** are applicable for chain formers. Deviations are anticipated if systems contain a significant number of rings or other non-chain structures.

## 3. Materials and Methods.

### 3.1 Materials

Five MAs were investigated: 2-ethyl-1-hexanol (2E1H) (Sigma Aldrich, Product No. 538051), 5-methyl-2-hexanol (5M2H) (Sigma Aldrich, Product No. 189731), 3,7-dimethyl-1-octanol (3,7D1O) (Fischer Scientific, CAS No. 106-21-8), 1-butanol (1-BL), and 1-propanol (1-PL). The results of 1-butanol (1-BL) and 1-propanol (1-PL) are from previously published results.[53, 72] For 2E1H, 5M2H, and 3,7D1O, they were kept in the dried molecular sieves (Sigma Aldrich, Product No. 208574) for 24 hours prior to measurements. The chemical structures of these MAs are shown in **Figure 3**. According to previous studies[1, 6, 16, 17, 53, 72-74] and as shown in the results section, these MAs form supramolecular chain structures with pronounced Debye peak.

OH
2E1H
OH
5M2H
OH
3,7D1O
OH
1-BL
OH
1-PL

**Figure 3.** Molecular structures of 2-ethyl-1-hexanol (2E1H), 5-methyl-2-hexanol (5M2H), 3,7-dimethyl-1-octanol (3,7D1O), 1-butanol (1-BL), and 1-propanol (1-PL). All monohydroxy alcohols (MAs) form supramolecular chain structures.

### 3.2 Broadband Dielectric Spectroscopy (BDS)

Broadband dielectric spectroscopy (BDS) was employed to quantify the dynamics of 2E1H, 5M2H, and 3,7D1O. MAs were sandwiched by two gold-plated electrodes with a diameter of 20 mm. A hollow Teflon spacer of a thickness 0.14 mm, an inner diameter of 16 mm, and an outer diameter of 25 mm was used between the electrodes to prevent electric shortage. The sandwiched samples were loaded onto a ZGS sample holder of Novocontrol Concept-40 system with an Alpha-A impedance analyzer and a Quatro Cryosystem temperature controller (temperature accuracy of $\pm 0.1$ K). All measurements covered angular frequencies from $6.28 \times 10^{7}$ to $6.28 \times 10^{-2}$ rad/s, and were with a root-mean-squared AC voltage of 1.0 V. The measurements were conducted from 293 K to 183 K at an interval of 10 K and from 183 K to 153 K at an interval of 5 K. A thermal annealing of 20 minutes was applied before each measurement to assure the achievement of thermal equilibrium.

We analyze the dielectric spectra in the following ways: **(i)** the Havriliak-Negami (HN) function fit; **(ii)** the relaxation time distribution analysis; and **(iii)** the LCM analysis. The LCM analysis follows **Eqns. 5-14**. In the HN-function fit, the complex dielectric permittivity, $\varepsilon^{*}(\omega)$, can be described through a summation of several HN functions:[75]

$$\varepsilon^{*}(\omega) = \varepsilon'(\omega) - i\varepsilon''(\omega) = \sum_{j} \frac{\Delta\varepsilon_{j}}{(1 + (i\omega\tau_{HN,j})^{\beta_{j}})^{\gamma_{j}}} + \varepsilon_{\infty} \quad (17)$$

where $\tau_{HN,j}$, $\beta_{j}$, and $\gamma_{j}$ are the HN time and the shape parameters of the $j^{th}$ dielectric relaxation. The average relaxation time of the $j^{th}$ process can be estimated as $\tau_{j} = \tau_{HN,j}\left[sin\left(\frac{\beta_{j}\pi}{2+2\gamma_{j}}\right)\right]^{-1/\beta_{j}}\left[sin\left(\frac{\beta_{j}\gamma_{j}\pi}{2+2\gamma_{j}}\right)\right]^{1/\beta_{j}}$. Note that the same set of fit parameters were used to fit the storage permittivity, $\varepsilon'(\omega)$, loss permittivity spectra, $\varepsilon''(\omega)$, and the derivative spectra, $\varepsilon'_{der}(\omega) = -\frac{\pi}{2}\frac{\partial \varepsilon'(\omega)}{\partial ln\omega}$.

In the relaxation time distribution analysis, the complex dielectric function is described through a superposition of Debye functions:

$$\varepsilon^*(\omega) - \varepsilon_\infty = \Delta\varepsilon \int_{-\infty}^{+\infty} \frac{g(ln\tau)}{1 + i\omega\tau} dln\tau \qquad (18)$$

where $g(ln\tau)$ represents the relaxation time distribution density function and a normalization condition $\int_{-\infty}^{+\infty} g(ln\tau)dln\tau = 1$ holds. A generalized regularization method (GENEREG) was employed to solve the integral equation for $g(ln\tau)$,[76] which have been successfully applied to the analyses of dielectric spectra previously.[17, 40, 77-79] In **Eqn. 18**, we assume Debye relaxation as the elementary mode in the relaxation time distribution analysis, due to the presence of a strong Debye process. The purpose of the relaxation time distribution analysis is to separate the overlapping relaxation processes rather than quantitative analyses of the dielectric relaxation amplitudes. More involved analysis has been pursued that uses other functional forms such as a HN function in the relaxation time distribution analysis,[80] which we do not pursue in this work.

### 3.3 Rheology

Linear rheology was applied to characterize the viscoelastic properties of MAs. Small amplitude oscillatory shear (SAOS) measurements were conducted on an Anton Paar (MCR302) rheometer from their glass transition temperatures, $T_g$, to temperatures where the terminal modes were resolved. The temperature was controlled by a CTD600 oven equipped with a liquid nitrogen evaporation unit (EVU20) with an accuracy of $\pm 0.1$ K. Following a previous protocol,[40, 81] we use parallel plates of 4 mm in diameter and vary the strain amplitude from $0.01\%$ near $T_g$ to $\sim 5\%$ in the terminal regime. All samples were loaded at room temperature and were cooled immediately to close to $T_g$ for measurements. During the measurements, the gap was maintained at $\sim 1\ mm$ and the angular frequency was set as $628 - 0.1$ rad/s. A thermal annealing of $20$ minutes was applied before each measurement to assure thermal equilibrium. The linear viscoelastic master curves of these MAs were obtained through applying the time-temperature superposition principle. The rheological structural relaxation time, $\tau_\alpha^R \approx 1/\omega_\alpha$ where $\omega_\alpha$ is the crossover between the storage moduli $G'(\omega)$ and $G''(\omega)$ at high frequencies, and the flow time $\tau_f \approx 1/\omega_f$ where $\omega_f$ is the onset of $G''(\omega) \sim \omega^2$ that matches well with the onset of the plateau of the amplitude of the complex viscosity, $|\eta^*(\omega)| = \sqrt{G'^2 + G''^2}/\omega$, in the zero-rate limit.[82]

## 4. Results and Discussions

### 4.1 Broadband dielectric spectroscopy.

**Figure a-4c** and **Figure a-5c** provide the dielectric loss permittivity, $\varepsilon''(\omega)$, and the derivative spectra, $\varepsilon'_{der}(\omega)$, of 2E1H **(Figures 4a** and **5a)**, 5M2H **(Figures 4b** and **5b)**, 3,7D1O **(Figures 4c** and **5c)**. **Figures 4d** and **4e** present $\varepsilon''(\omega)$ of 1-BL and 1-PL from Ref. 53 and Ref. 72. A pronounced Debye relaxation with time, $\tau_D$, at low frequencies and a structural relaxation with time, $\tau_\alpha^{BDS}$, at high frequencies are well resolved. The superscript *BDS* indicates results from BDS measurements. Recent studies suggest relaxation processes at intermediate times, $\tau_m^{BDS}$, between $\tau_\alpha^{BDS}$ and $\tau_D$.[10, 17] In addition,

one can observe a secondary relaxation with relaxation time, $\tau_\beta$, at frequencies higher than the structural relaxation. Representatives of these relaxation processes are pointed out by arrows in each figure panel. These observations agree well with previous BDS measurements of these MAs.[1, 6, 16, 17, 74]

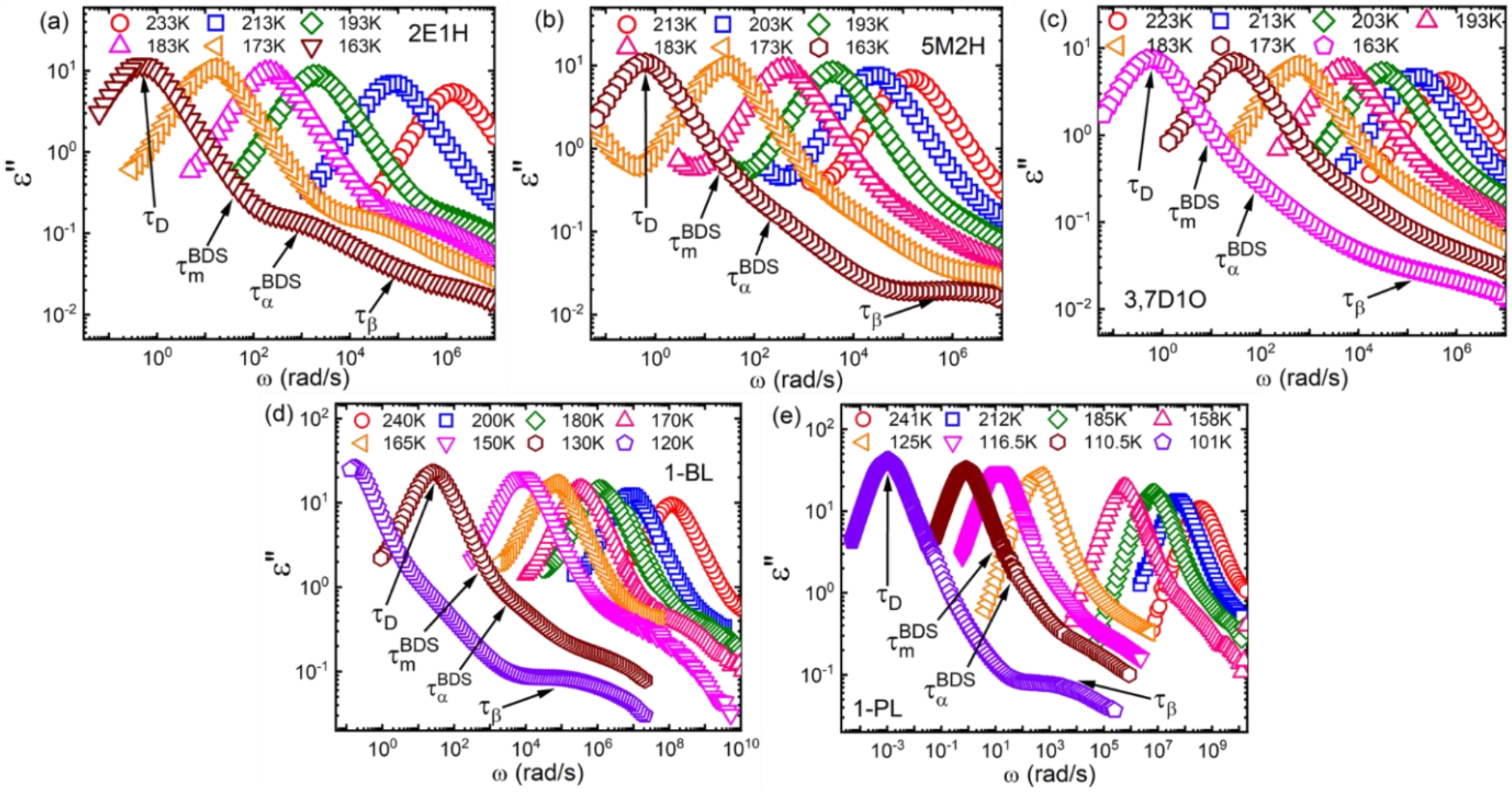

**Figure 4**. Dielectric loss permittivity $\varepsilon''(\omega)$ of **(a)** 2E1H; **(b)** 5M2H; **(c)** 3,7D1O; **(d)** 1-BL; and **(e)** 1-PL. The arrows point to the corresponding molecular processes, including the Debye process with relaxation time, $\tau_D$, the intermediate process with relaxation time, $\tau_m^{BDS}$, the structural relaxation with relaxation time, $\tau_\alpha^{BDS}$, and the secondary relaxation with relaxation time, $\tau_\beta$.

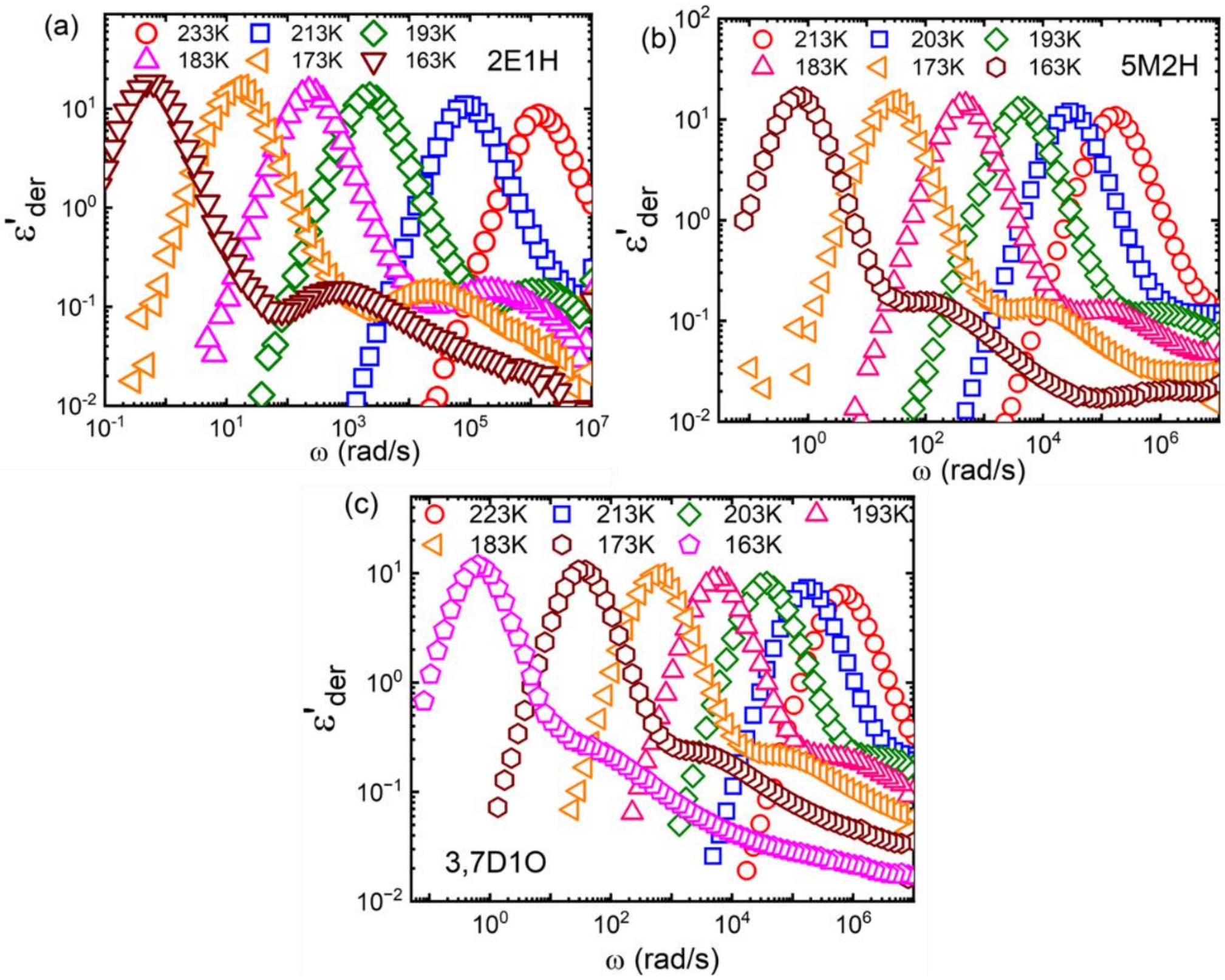

**Figure 5.** Derivative spectra $\varepsilon'_{der}(\omega) = -\frac{\pi}{2}\frac{\partial \varepsilon'}{\partial(\ln \omega)}$ of **(a)** 2E1H; **(b)** 5M2H; and **(c)** 3,7D1O at representative temperatures. The Debye relaxation and the structural relaxation are well separately and well resolved in each case.

**Figure 6** provides the representative HN-function fit to $\varepsilon''(\omega)$ (**Figures 6a**, **6e**, **6i**, **6m**, and **6p**) and $\varepsilon'_{der}(\omega)$ (**Figures 6b**, **6f**, **6j**, and **6n**) of each MAs. The dashed, dotted, and the dash-dotted lines represent the Debye process with relaxation time $\tau_D$, the intermediate process with relaxation time $\tau_m^{BDS}$, and the structural relaxation with relaxation time $\tau_\alpha^{BDS}$, respectively. The solid black lines are the sum of these HN functions. As pointed out in a previous study,[17] two processes (a Debye process and a structural relaxation process) cannot fit well $\varepsilon''$ and $\varepsilon'_{der}$ simultaneously for 2E1H, 5M2H, and 1-BL. The same is observed for 3,7D1O (see **Figure S1** of the **Supplementary Materials (SM)**). **Figures 6c, 6g, 6k, 6o, 6q** provide the relaxation time distribution density function, $\Delta\varepsilon \times g(ln\tau)$, of the corresponding dielectric spectra, where an intermediate relaxation process is well-resolved.

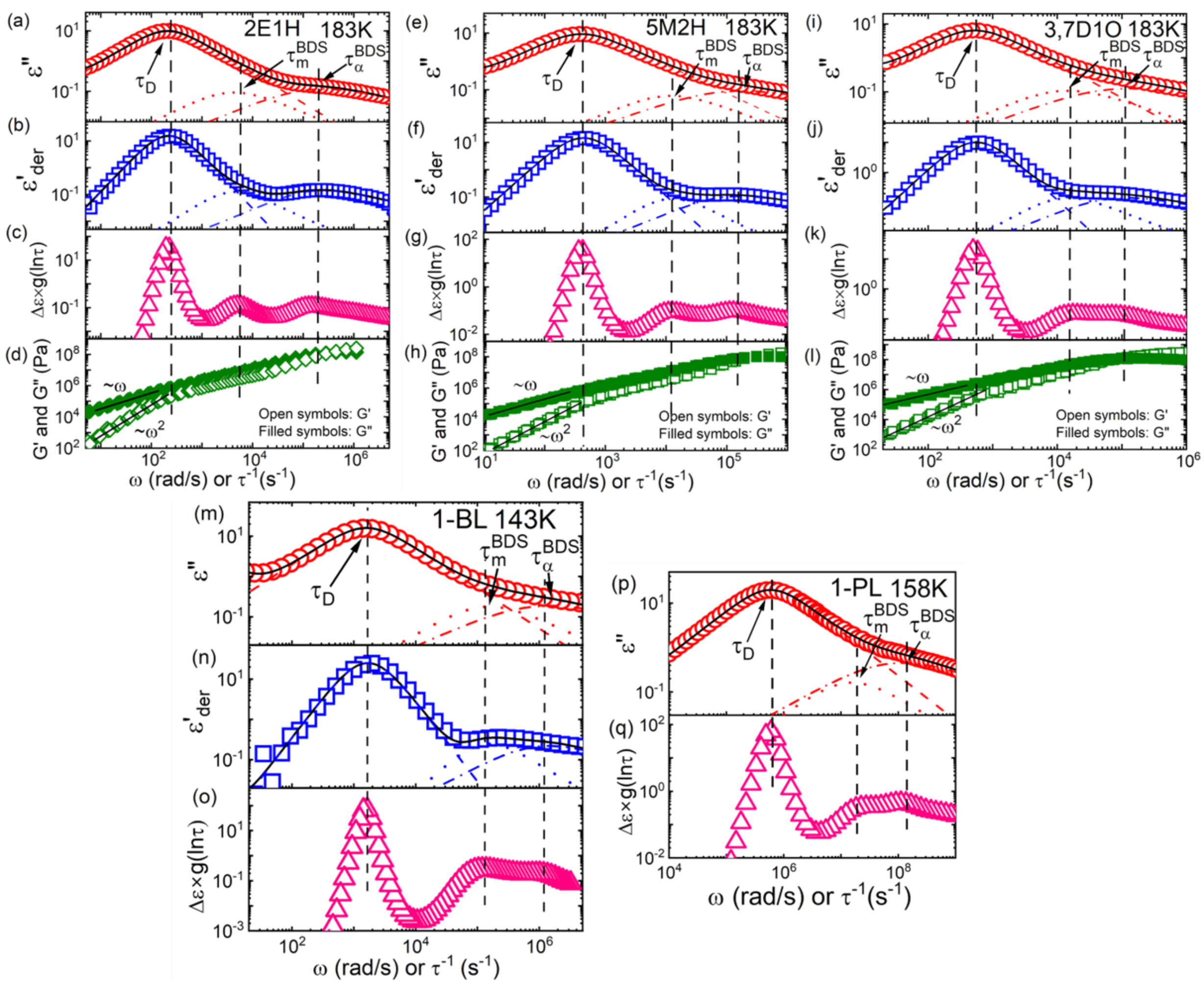

**Figure 6**. Comparisons of the Havriliak-Negami (HN) function analysis and the relaxation time distribution analysis to 2E1H (**6a – 6c**), 5M2H (**6e - 6g**), 3,7D1O (**6i - 6k**), 1-BL (**6m - 6o**), and 1-PL (**6p - 6q**). In addition, the linear viscoelastic master curves of 2E1H (**6d**), 5M2H (**6h**), and 3,7D1O (**6l**) are also presented for comparison. The structural relaxation from the BDS measurements agrees well with the characteristic crossover frequency between storage and loss modulus. The characteristic angular frequencies of Debye relaxation match well with the onset of the terminal relaxation in each MA. The relaxation time distribution analysis demonstrates the presence of an intermediate relaxation process between the structural relaxation and the Debye relaxation (also see the enlarged figure of them in **Section 4.2** below).

**Figure 7** summarizes $\tau_D$ (red open circles), $\tau_m^{BDS}$ (blue open squares), $\tau_\alpha^{BDS}$ (olive open diamonds), and $\tau_\beta$ (black open upper-triangles) of MAs from HN-function fit. The dielectric relaxation time faster than $10^{-7}$ s of 2E1H, 5M2H, and 3,7D1O are from Ref. 3 and Ref. 73. The results of 1-BL are from Ref. 1 and 1-PL are from Ref. 83. The characteristics of $\tau_{\alpha}^{BDS}$ and $\tau_D$ of these MAs have been discussed in the past,[6] which include a Vogel-Fulcher-Tamann (VFT) temperature dependence of at $\sim T_g$ and $\sim T_g + 100\ K$, and a kink in the activation plot of $\tau_D$ at $\tau_D \sim 10^{-9} - 10^{-8}\ s$.[73] The kink at $\tau_D \sim 10^{-9} - 10^{-8}\ s$ has been viewed as a dynamic anomaly. In addition, $\tau_{m}^{BDS}$ shows an Arrhenius-like temperature dependence at intermediate temperatures with deviations both at the high and low temperatures. $\tau_{\beta}$ is a secondary relaxation following Arrhenius temperature dependence with apparent activation energy ranging from ~56 kJ/mol to ~28 kJ/mol depending on the chemical structures of the MAs. Furthermore, we summarize the shape parameters, $\beta$ (open symbols) and $\gamma$ (filled symbols), of all three processes from the HN function analysis in **Figure 8**, and the dielectric relaxation strength of the Debye process, $\Delta\varepsilon_D$, the intermediate process, $\Delta\varepsilon_m$, and the structural relaxation, $\Delta\varepsilon_{\alpha}$, in **Figure 9**. As shown in **Figure 8**, all Debye processes give $\beta = \gamma = 1.0$, as expected. The intermediate relaxation processes have $\beta \approx 0.8 - 0.9$ and $\gamma \approx 0.8 - 0.9$. The structural relaxation processes have shape parameters of $\beta \approx 0.7 - 0.8$ and $\gamma \approx 0.5 - 0.6$, which are close to the recently revealed generic shapes of the structural relaxation of molecular liquids.[41] $\Delta\varepsilon_D$ and $\Delta\varepsilon_{\alpha}$ increase with cooling. Interestingly, $\Delta\varepsilon_B$ is smaller than $\Delta\varepsilon_D$ and $\Delta\varepsilon_{\alpha}$. In addition, $\Delta\varepsilon_B$ has a much weaker temperature dependence and can either increase or decrease with temperature.

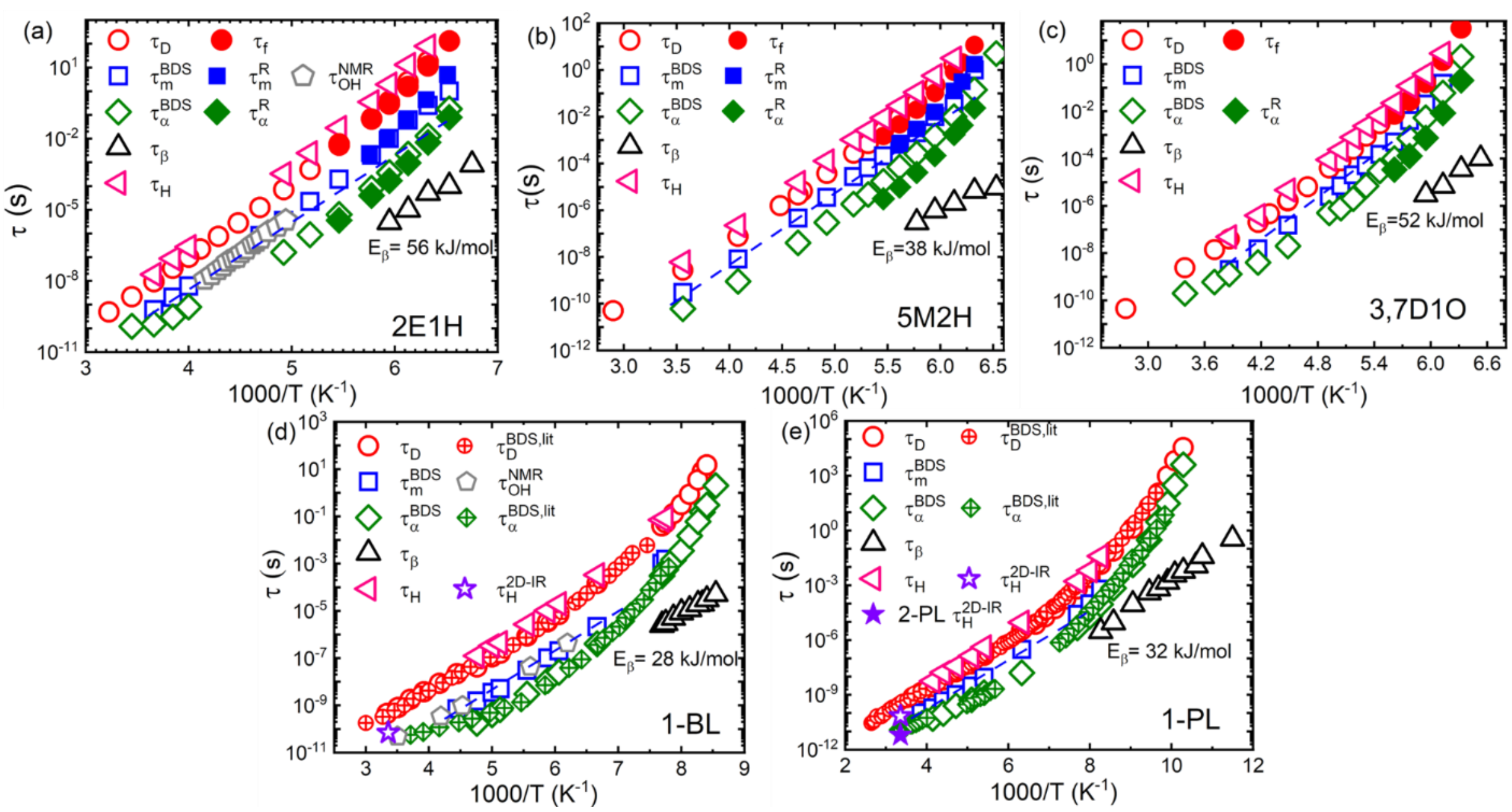

**Figure 7** Temperature dependence of $\tau_{\mathrm{D}}$ (red open circles), $\tau_m^{BDS}$(blue open squares), $\tau_\alpha^{BDS}$ (olive open diamonds), and $\tau_\beta$ (black open upper triangles) of broadband dielectric spectroscopy measurements of **(a)** 2E1H, **(b)** 5M2H, **(c)** 3,7D1O, **(d)** 1-BL, and **(e)** 1-PL. The red filled circles, blue filled squares, and olive filled diamonds of panels (a), (b), and (c) are the corresponding terminal relaxation time, the intermediate times, and the structural relaxation from linear viscoelastic measurements for 2E1H, 5M2H, and 3,7D1O. The red circles filled with a cross and olive diamonds filled with a cross of 1-BL and 1-PL are from literature dielectric measurements with Ref. 34 (for 1-BL) and Ref. 72 (for 1-PL). The grey open pentagons of panels **(a)** and **(d)** are the hydrogen bonding exchange time from NMR measurements from the results of Ref.34 (for 2E1H) and Ref.1 (for 1-BL). The pink left triangles represent the calculations for the H-bonding lifetime, $\tau_H$, from the living chain model (LCM). The dashed blue lines are guides to eye to show the Arrhenius nature of $\tau_m^{BDS}$ at intermediate temperatures. The purple open stars in panels **(d)** and **(e)** are results of 2D-IR measurements or femtosecond IR measurements of 1-BL and 1-PL for the hydrogen bonding lifetime from Refs. 86 and 87. The purple filled star in panel **(e)** is a 2D-IR measurements of 2-PL from Ref. 85.

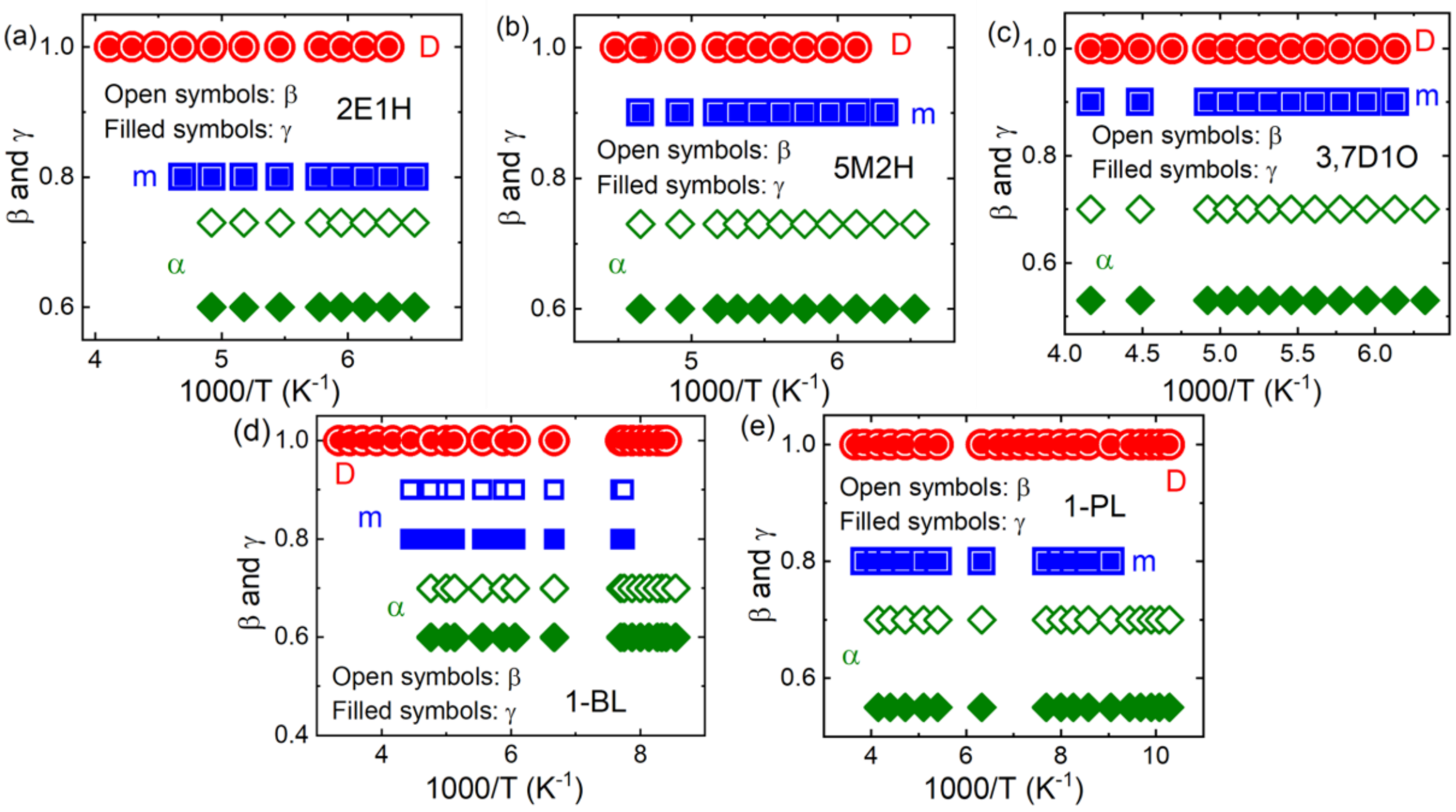

**Figure 8.** Shape parameters, $\beta$ (open symbols) and $\gamma$ (filled symbols), of the HN-function fit to the Debye process (red symbols), the intermediate process (blue symbols), and the structural relaxation (olive symbols) of **(a)** 2E1H, **(b)** 5M2H, **(c)** 3,7D1O, **(d)** 1-BL, and **(e)** 1-PL. $D$ represents the Debye process, $m$ represents the intermediate process, and $\alpha$ represents the structural relaxation.

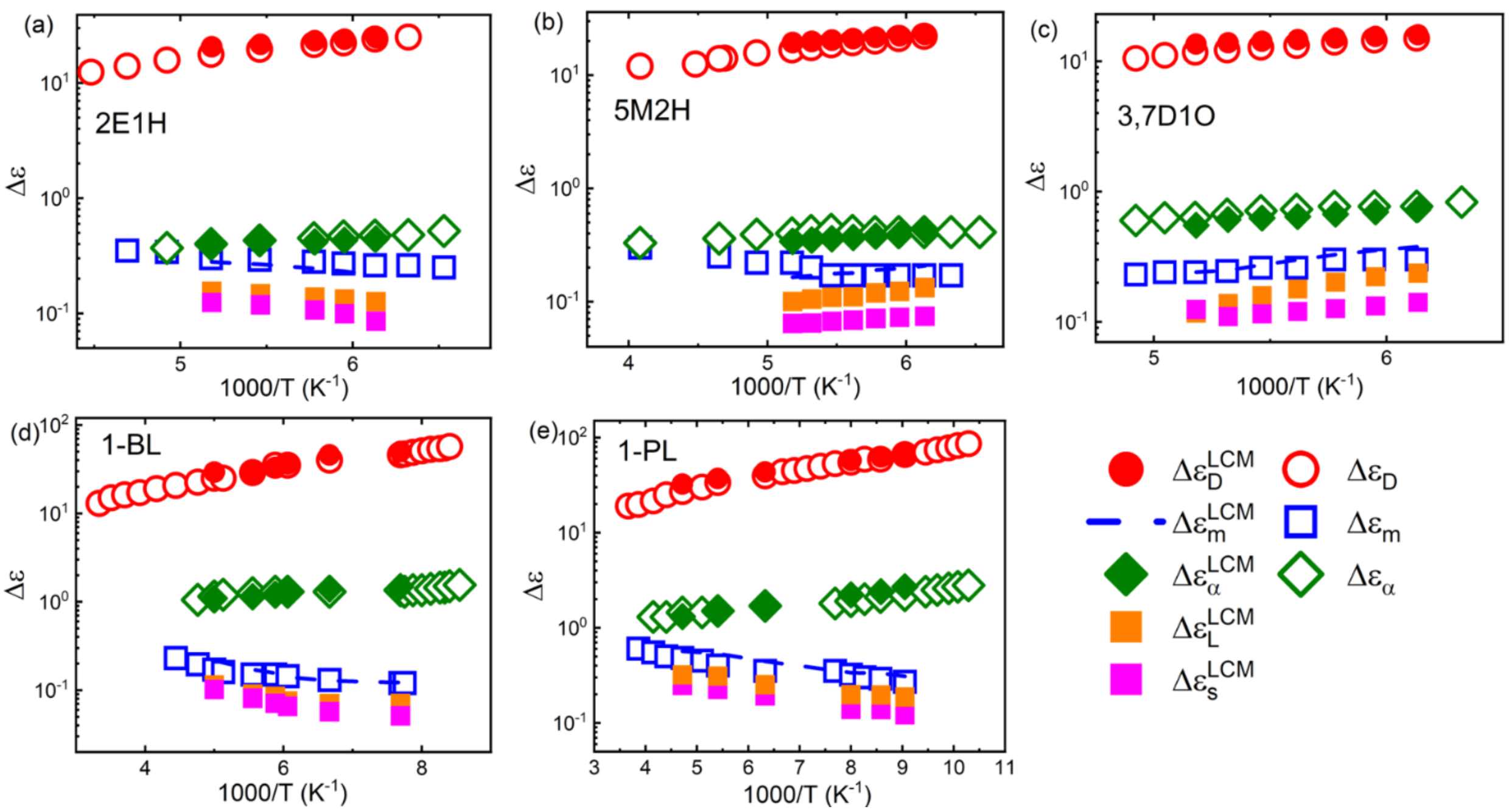

**Figure 9**. Dielectric relaxation amplitude, $\Delta\varepsilon$, from the HN-function fit analysis of the Debye process $\Delta\varepsilon_D$ (red open circles), the intermediate process $\Delta\varepsilon_m$ (blue open squares), and the structural relaxation $\Delta\varepsilon_\alpha$ (olive open diamonds) of **(a)** 2E1H, **(b)** 5M2H, **(c)** 3,7D1O, **(d)** 1-BL, and **(e)** 1-PL. The dielectric relaxation amplitudes of each process from LCM analysis have also been presented for comparison, including the Debye process, $\Delta\varepsilon_D^{LCM}$ (the filled red circles), the chains longer than $N_R$, $\Delta\varepsilon_L^{LCM}$ (the filled orange squares), the chains equal to or shorter than $N_R$, $\Delta\varepsilon_S^{LCM}$ (the filled pink squares), and the structural relaxation, $\Delta\varepsilon_\alpha^{LCM}$ (the filled olive diamonds). Excellent agreements have been observed between the HN function fit and the LCM analysis for all relaxation processes. Note that the blue dashed lines represent the sum of $\Delta\varepsilon_L^{LCM}$ and $\Delta\varepsilon_S^{LCM}$ that agree well with $\Delta\varepsilon_m$ from the HN analyses.

**4.2 Linear rheology.**

Error! Reference source not found. **6h,** and **6l** provide the linear viscoelastic master curves of 2E1H, 5M2H, and 3,7D1O. The enlarged linear viscoelastic master curves of these three MAs are given in **Figure 10**. The van Gurp-Palmen plots[84] of phase angle $\delta$ vs the amplitude of the complex modulus, $|G^*|$, are presented in **Figure S2** of the Supplementary Materials **(SM)**. Rheological simplicity holds only at a limited temperature ranges close to $T_g$ for 2E1H, 5M2H, and 3,7D1O, along with noticeable deviation for all three MAs at high temperatures. This indicates a potential breakdown of the time-temperature superposition. As discussed in a following section, the breakdown of the time-temperature superposition of MAs correlate well with a sharp change of the supramolecular chain length, highlighting the large influence of the supramolecular structure formation on the dynamics of MAs. The corresponding dynamics shift factors, $a_T$, are presented in the inset of the corresponding panels of **Figure S2** of the **SM**. No vertical shifts are needed for the construction of the master curves. In all MAs, $G' \sim \omega^2$, and $G'' \sim \omega$ (the solid lines of **Figures 6d, 6h,** and **6l**) have been observed at high temperatures, supporting the access of terminal flow.

From the linear viscoelastic spectra, one can identify the characteristic structural relaxation time, $\tau_\alpha^R$, from the high-frequency peak of the loss moduli, $G''(\omega)$, with $\tau_\alpha^R \approx$ $1/\omega_\alpha$. Here, the superscript $R$ implies results from rheological measurements. The terminal relaxation, $\tau_f$, can be estimated from the onset frequency where the dynamics viscosity deviates from the zero-shear limit or the corresponding frequency of the scaling of $G' \sim \omega^2$ with $\tau_f \approx 1/\omega_f$ (**Figure 10**). As shown in **Figure 6** the direct comparison between the BDS spectra and the linear viscoelastic spectra as well as the temperature dependence of dynamics of **Figure 7**, $\tau_\alpha^R \approx \tau_\alpha^{BDS}$ and $\tau_f \approx \tau_D$ hold for these MAs.

A close look at the viscoelastic master curves of 2E1H, 5M2H, and 3,7D1O of the enlarged viscoelastic spectra (**Figure 10**) reveal another interesting feature: there is a dynamics scaling change of the storage modulus, $G'(\omega)$, at intermediate frequencies: $G'(\omega) \sim \omega$ before $G'(\omega) \sim \omega^2$ upon reducing $\omega$. These observations indicate the presence of intermediate relaxation times. Interestingly, one can identify the intermediate times, $\tau_m^R$, from the LCM analysis (the dashed lines, dash-dotted lines, and dash-dot-dotted lines in **Figure 10**) that agrees well with $\tau_m^{BDS}$, $\tau_m^R \approx \tau_m^{BDS}$ (see **Figures 6** and **7**). Furthermore, $\tau_m^{BDS}$ at high temperatures matches well with the H-bonding exchange time from NMR measurements (**Figure 7**), indicating the H-bonding breakage and recombination as the origin of the intermediate process.[17] Therefore, $\tau_m^R \approx \tau_m^{BDS} \approx \tau_B$, supporting the LCM's description of the supramolecular chain breakage and chain swapping process at intermediate times.[16, 17]

The relaxation process at $\tau_m^R$ from linear viscoelastic measurements can be rationalized from energy dissipation associated with the chain breakage and swapping. It is worth noting that the chain breakage and swapping do not lead to large dipole moment change.[16] The emergence of dielectric relaxation process is thus not directly associated with the dipole moment change of the chain breakage and swapping process. Instead, the chain breakage truncates the relaxation time distribution of the normal modes at $\tau_m^{BDS}$. The intermediate relaxation processes of $\tau_m^{BDS}$ are thus the normal modes with relaxation times slower than or equal to $\tau_m^{BDS}$. These results are consistent with the recently observed scaling behaviors of the stress relaxation, $G(t)$, of MAs with strong chain formations, where $G(t) \sim t^{-1/2}$ switches to $G(t) \sim t^{-1}$ at intermediate times.[17]

**4.3 The living chain model (LCM) analysis.**

The above analysis provides clear characterizations of $\tau_\alpha$, $\tau_m$ (or $\tau_B$), and $\tau_D$ (or $\tau_f$). According to LCM, $N_R \approx \sqrt{\tau_B/\tau_\alpha}$ and $\bar{N} \approx (\tau_D^2/(\tau_B\tau_\alpha))^{1/2}$. The volume fraction of supramolecular chain with length $i$, $\varphi(i)$, can be obtained from **Eqn. 4**. The solid lines of **Figure 10** are the LCM prediction and their comparison with the linear viscoelastic master curves. The dashed lines, dash-dotted lines, and dash-dot-dotted lines of **Figure 10** provide the detailed contribution of $G_\alpha^*(\omega)$, $G_S^*(\omega)$, and $G_L^*(\omega)$. In addition, **Figure 11** offers the LCM predictions (lines) on the dielectric spectra of these MAs and their comparison with experiments (symbols), where the free parameters are $C_{\bar{N}} g_{\bar{N}}$, $g_\alpha$, and $\epsilon$. The dashed blue lines, the dash-dot-dotted orange lines, the dash-dotted pink lines, and the dashed green lines of **Figure 11** represent the specific contributions of $\varepsilon_D^*(\omega)$, $\varepsilon_L^*(\omega)$, $\varepsilon_S^*(\omega)$, and $\varepsilon_\alpha^*(\omega)$. The solid lines in **Figure 11** show the sum of the contributions of all four processes. The comparisons between LCM and BDS spectra at different temperatures are also given in **Figure S3** of the Supplementary Materials (**SM)**. As shown in **Figures 10** and **11**, excellent agreements are found between experiments and the LCM's prediction on the dielectric properties and rheological properties of MAs, highlighting the applicability of the LCM to describe the structures and dynamics of MAs.

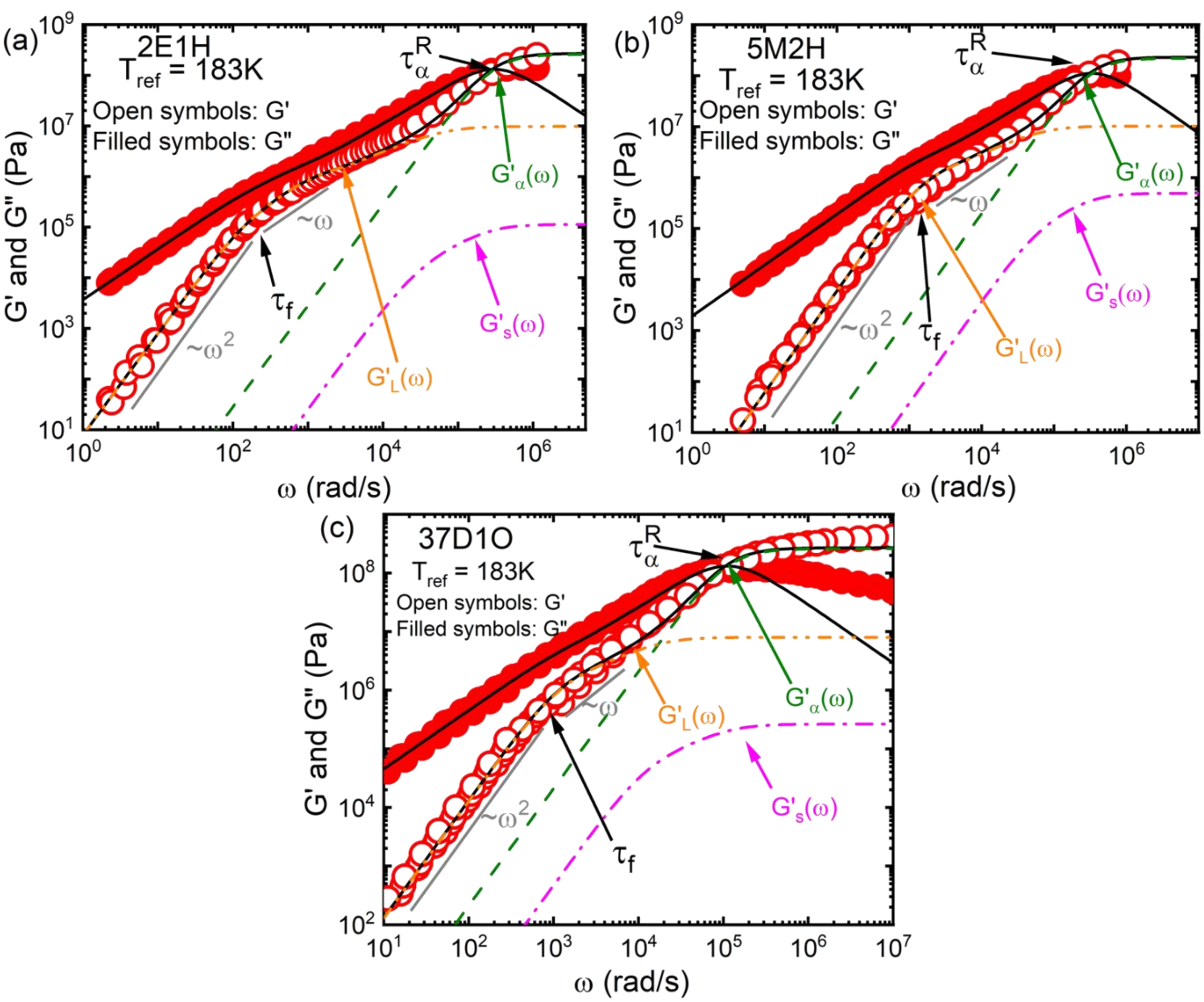

**Figure 10.** Comparison of the LCM and linear viscoelastic master curves of **(a)** 2E1H, **(b)** 5M2H, and **(c)** 3,7D1O at reference temperature of $T = 183\ K$. The dashed green lines, the dash-dotted pink lines, the dash-dot-dotted orange lines represent the $G'_\alpha(\omega)$, $G'_S(\omega)$, and $G'_L(\omega)$. The solid black lines are the sum of all three contributions, where excellent agreements between experiments and LCM predictions are observed.

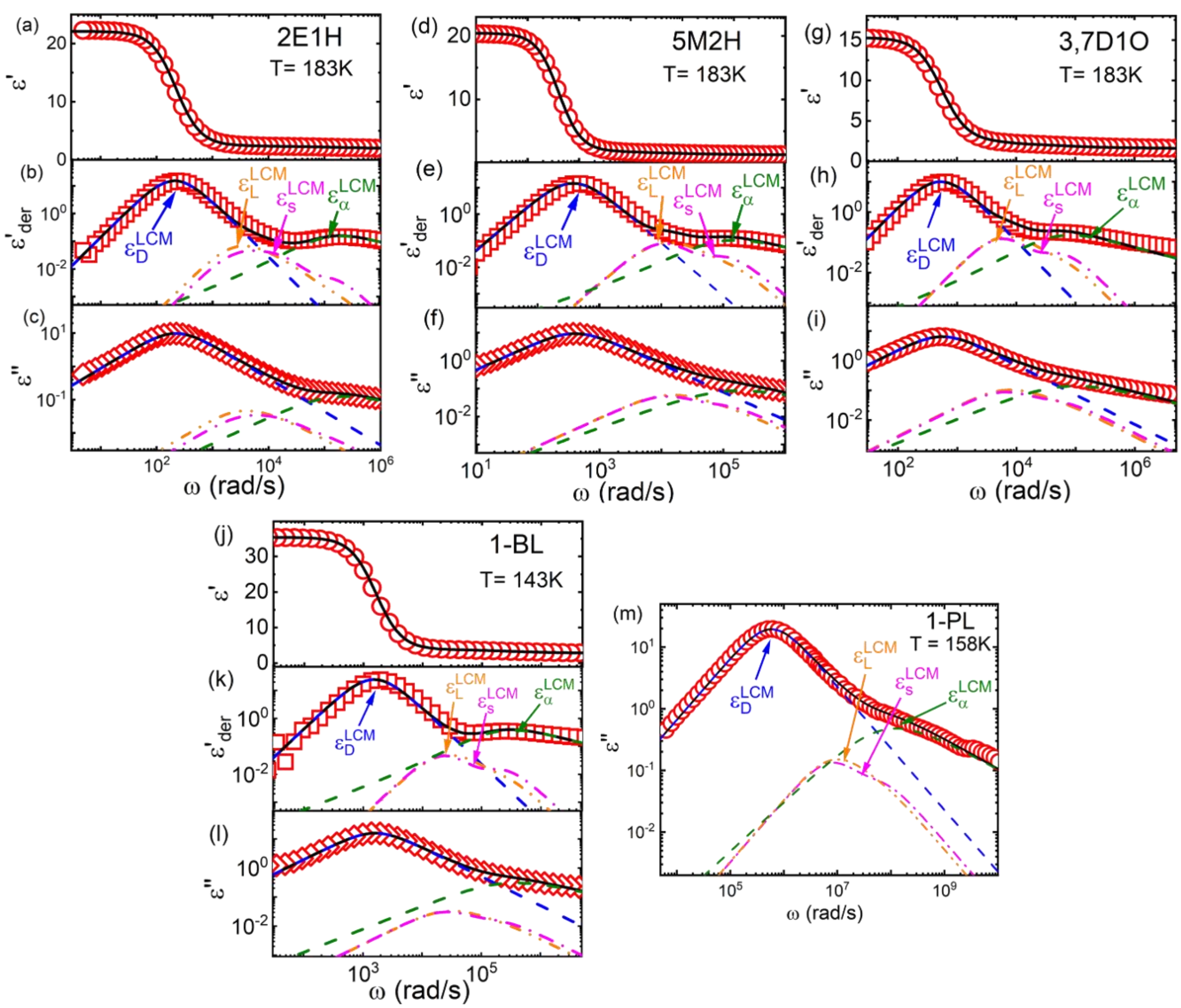

**Figure 11.** Comparison of LCM and the dielectric spectra of **(a)** 2E1H, **(b)** 5M2H, **(c)** 3,7D1O, **(d)** 1-BL, and **(e)** 1-PL at corresponding reference temperatures labelled in each panel. The dashed green lines, the dash-dot-dotted orange lines, the dash-dotted pink lines, and the dashed blue lines represent $\varepsilon_\alpha^*(\omega)$, $\varepsilon_L^*(\omega)$, $\varepsilon_S^*(\omega)$, and $\varepsilon_D^*(\omega)$. The solid black lines are the sum of the four dielectric processes. Excellent agreement between LCM and experiments is observed in all MAs.

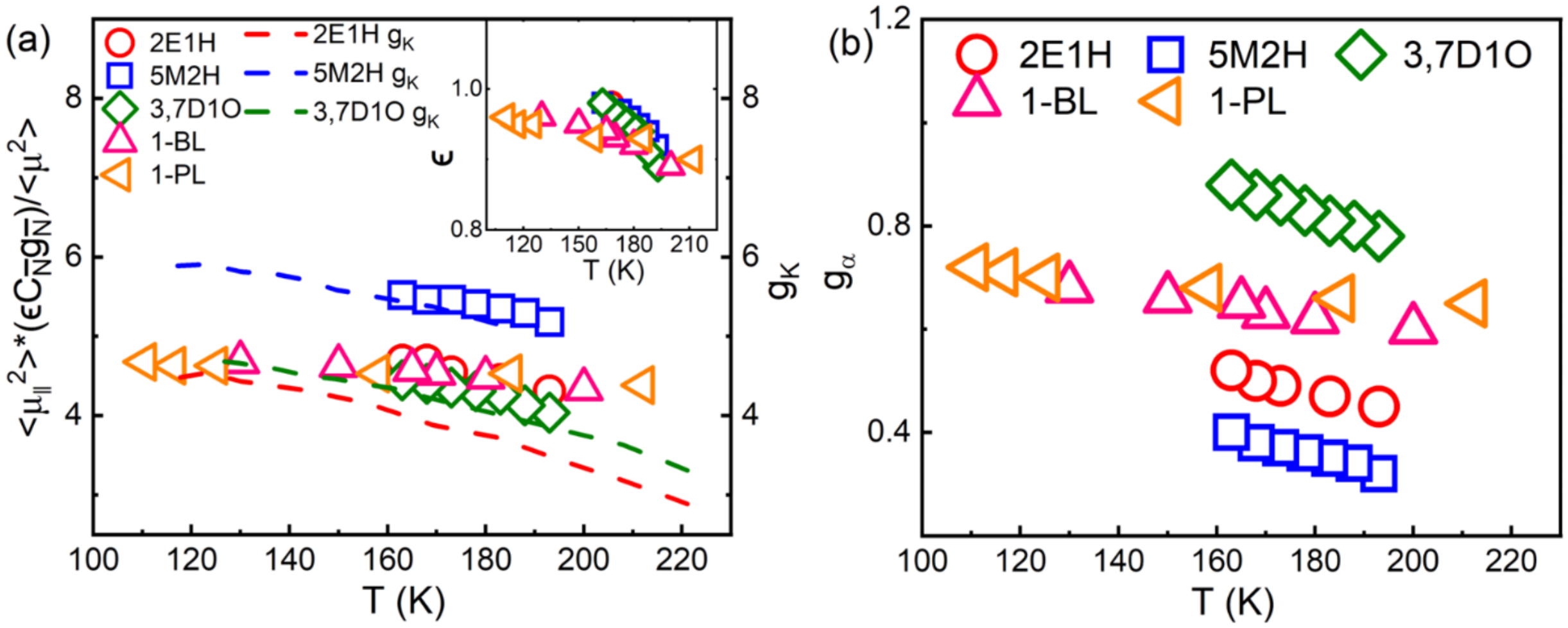

**Figure 12 (a)** $\langle\mu_{||}^{2}\rangle\epsilon C_{\bar{N}}g_{\bar{N}}/\langle\mu^{2}\rangle$ (symbols), and **(b)** $g_{\alpha}$ of the LCM analyses. The Kirkwood-Fröhlich factor, $g_{k}$ (dashed lines), of MAs is obtained from experiments through **Eqn.19.** The inset of the **(a)** gives $\epsilon$ at different temperatures.

**Figure 12** provides $\langle\mu_{||}^{2}\rangle\epsilon C_{\bar{N}}g_{\bar{N}}/\langle\mu^{2}\rangle$ and $g_{\alpha}$ and their comparison to $g_{k}$ values obtained from the BDS measurements, where $\langle\mu_{||}^{2}\rangle/\langle\mu^{2}\rangle = 0.80$ is a constant. Several features are worth noting: **(i)** $\langle\mu_{||}^{2}\rangle\epsilon C_{\bar{N}}g_{\bar{N}}/\langle\mu^{2}\rangle \sim 4-6$ is observed, and is comparable to $g_{K}$ values from BDS measurements. **(ii)** $g_{\alpha}\sim 0.3-1$ is observed and changes only slightly with temperature. Given the simplicity of the assumptions underlying the calculation of the dielectric amplitude associated with the structural relaxation, the resulting value of $g_{\alpha}$, being on the order of unity, is quite reasonable. **(iii)** At a given temperature, the following order is observed of $\langle\mu_{||}^{2}\rangle\epsilon C_{\bar{N}}g_{\bar{N}}/\langle\mu^{2}\rangle$: 5M2H > 2E1H ~ 3,7D1O > 1-BL > 1-PL. Among these MAs, 5M2H is a secondary alcohol, which is expected to have a higher backbone rigidity than the rest. This might explain its high $\langle\mu_{||}^{2}\rangle\epsilon C_{\bar{N}}g_{\bar{N}}/\langle\mu^{2}\rangle$. Within the primary alcohols, the longer the alkyl group, the larger the $\langle\mu_{||}^{2}\rangle\epsilon C_{\bar{N}}g_{\bar{N}}/\langle\mu^{2}\rangle$. **(iv)** $\langle\mu_{||}^{2}\rangle\epsilon C_{\bar{N}}g_{\bar{N}}/\langle\mu^{2}\rangle$ reduces with heating for all MAs. This is most likely due to a reduction of the volume fraction of the total supramolecular chains, $\epsilon$, as demonstrated in the inset of **Figure 12a**. These analyses thus further support the applicability of the LCM analyses for dielectric properties of MAs.

In addition, LCM can help decouple the dielectric relaxation strength of all relaxation processes. The filled symbols of **Figure 8** offer the dielectric relaxation amplitude of the Debye process, $\Delta\varepsilon_D^{LCM}$, $\Delta\varepsilon_L^{LCM}$, $\Delta\varepsilon_S^{LCM}$, and $\Delta\varepsilon_\alpha^{LCM}$. Note that both $\varepsilon_L^*(\omega)$ and $\varepsilon_S^*(\omega)$ contribute to the dielectric responses at intermediate frequencies, and $\Delta\varepsilon_L^{LCM} + \Delta\varepsilon_S^{LCM}$ (the blue lines) are comparable with $\Delta\varepsilon_m$ (the open blue squares). Excellent agreement between LCM calculation and experiments across a wide temperature range is thus obtained. Interestingly, $\Delta\varepsilon_S^{LCM}$ accounts a significant portion of $\Delta\varepsilon_m$ in all these MAs, although $\phi(N > N_R) \gg \phi(N \leq N_R)$ is observed. These analyses suggest that the *short* components ($N \leq N_R$) contribute actively to the dielectric relaxation at intermediate frequencies.

**4.4 The H-bonding lifetime, the supramolecular chain sizes, and the reaction rate constants.**

The excellent agreement between LCM prediction and the experiments inspires further discussion. For instance, $\tau_H \approx \tau_D N_R \approx \tau_D(\tau_B/\tau_\alpha)^{1/2}$, suggesting $\tau_H$ differs from the Debye relaxation time by a fact of $N_R$ that can be obtained directly from the dynamics measurements. At $N_R \geq 1$, $\tau_H \geq \tau_D$ holds, which indicates any supramolecular dynamics cannot survive longer than $\tau_H$. Conversely, an accurate measure of the H-bonding lifetime should last a time longer than $\tau_D$.

We note that 2D-IR or femtosecond IR[85-87] and NMR[1, 3, 34, 88] have been utilized to measure the H-bonding dynamics. For instance, the purple stars of **Figure 7d and 7e** give the H-bonding dynamics of 1-PL and 1-BL from these IR measurements.[85-87] The grey open pentagons of **Figures 7a and 7d** provides the H-bonding dynamics of 1-BL and 2E1H from NMR measurements.[1, 34] Interestingly, all these measurements (2D-IR and NMR) offer a time matching well with the H-bonding exchange time or the chain breakage and swapping time, $\tau_B$, from BDS. Given H-bonding lifetime should be at least no shorter than the longest time of the supramolecular structures, the Debye time, we believe these previous 2D-IR[85, 86] or femtosecond IR measurements[87] provide H-bonding exchange time, which differ from the H-bonding lifetime, $\tau_H$, by a factor of the supramolecular chain sizes, $\bar{N}$.

As shown in **Figure 7**, $\tau_H$ (pink left triangles) follows a near Arrhenius temperature dependence at high temperatures and switches to a super-Arrhenius temperature dependence close to $T_g$. This indicates a strong influence of the molecular cooperativity (i.e. configurational entropy) on chemical reactions (H-bonding association and dissociation) in the deep supercooled region. This is understandable since molecular diffusion in the supercooled region is strongly affected by the structural relaxation that requires molecular cooperativity. At high temperatures, $\tau_H \approx \tau_D$ is observed, indicating the H-bonding lifetime controls the Debye process at these temperatures. A deviation takes place between $\tau_H$ and $\tau_D$ at $N_R > 1$, which leads to a kink in temperature dependence of $\tau_D$ at $\sim 250\ K$ for many MAs.[73] Thus, LCM offers a natural molecular explanation for the previously observed 250 K anomaly in dynamics of MAs.

**Figure 13** summarizes $\bar{N}$ (mainframe) and $N_R$ (inset **Figure 13**) from the dielectric measurements. With the identification of $\tau_H$, one can also estimate $N_c \approx (\tau_H/\tau_\alpha)^{1/3}$, which are included as the red filled circles in **Figure 13**. Large supramolecular chain structures have been observed for these MAs, and $\bar{N} > N_c$ holds at all temperatures. For 5M2H, 3,7D1O, and 1-PL, $\bar{N}{\sim}10-60$ are observed over the temperature range of the measurements. For 2E1H, $\bar{N}$ can be as large as 200 in the deep supercooled region. In addition, 2E1H and 5M2H show an increment in chain length upon cooling. The chain lengths of 3,7D1O and 1-PL first increase and decrease upon cooling. The chain lengths of 1-BL decrease with cooling at the temperature range of the measurements. These observations imply a non-monotonic temperature dependence on the supramolecular chain length in these MAs with temperature. In contrast to the large variations in $\bar{N}$ with MA types and temperature, $N_R{\sim}1-5$ are found for all five MAs. In analogy to $\bar{N}$, $N_R$ also exhibits a non-monotonic temperature dependence with a peak temperature different from that of $\bar{N}$.

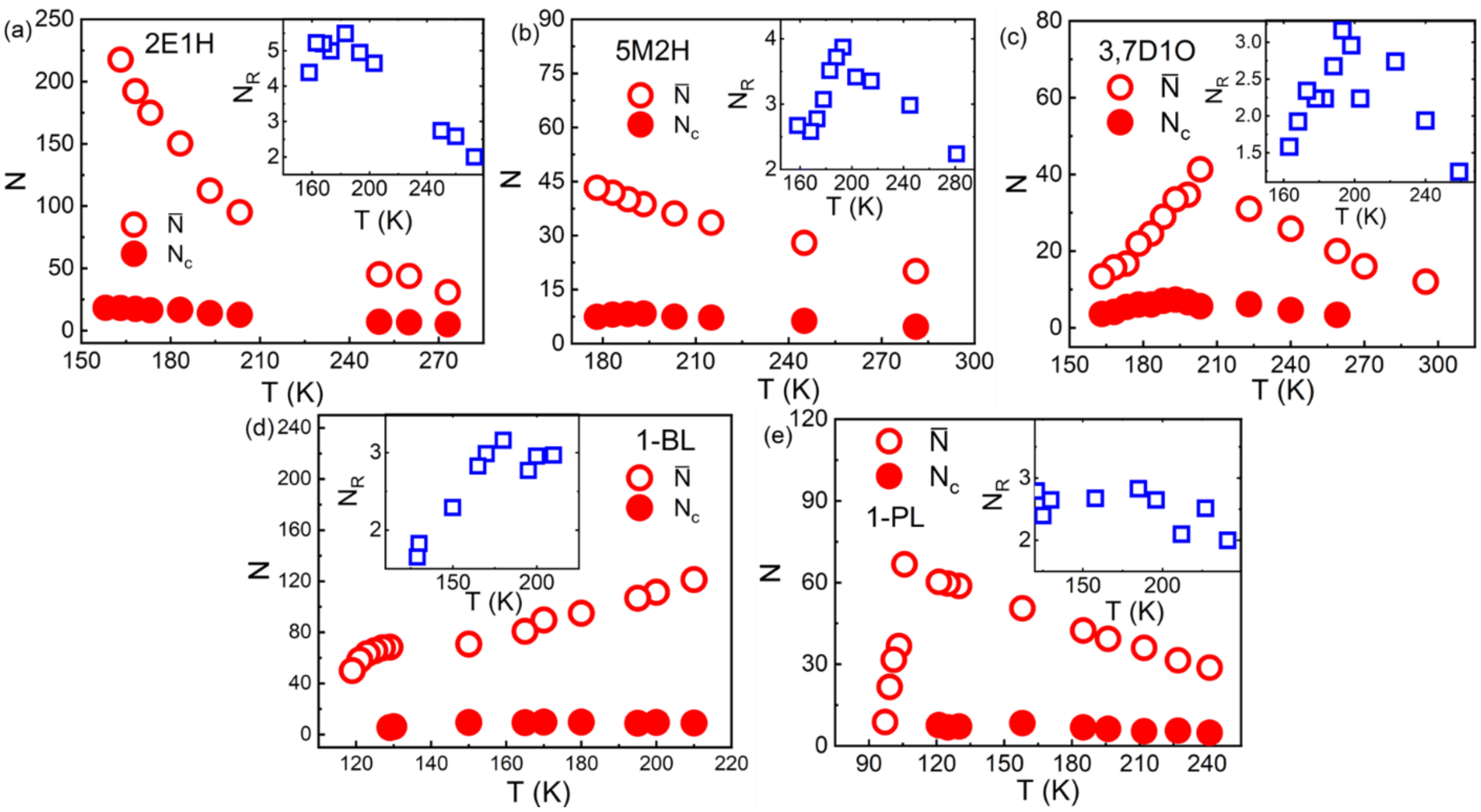

**Figure 13.** The average supramolecular chain length, $\bar{N}$ (the red open circles), the critical chain length, $N_c$ (the red filled circles), and the dynamic chain length, $N_R$ (the inset), of **(a)** 2E1H, **(b)** 5M2H, **(c)** 3,7D1O, **(d)** 1-BL, and **(e)** 1-PL.

Furthermore, the identification of $\tau_H$ provides a direct quantification of $k_2 = \frac{1}{\tau_H}$. At the same time, $\tau_D^2/\tau_\alpha = \left(\frac{c_0 k_1}{2k_2^3}\right)^{1/2}$ holds for a large number of MAs with different sizes and different types.[16] Thus, the dynamics measurements provide a direct estimate of $c_0 k_1$, where $c_0$ is the molar concentration of MA in the supramolecular chain structures. **Figure 14** provides $\ln(c_0 k_1)$ and $\ln(k_2)$ vs $1000/T$, where an Arrhenius temperature dependence has been observed for each MA over a wide temperature range. Deviations of the original Arrhenius temperature dependence of $\ln(c_0 k_1)$ and $\ln(k_2)$ have been observed at temperatures close to $T_g$, especially for 3,7D1O, 1-BL, and 1-PL. We attribute the deviation to an active interplay of glassy physics and chemical reaction kinetics. A direct estimate of the activation energy of the H-bonding association, $\Delta E_1$, and the H-bonding dissociation, $\Delta E_2$, can be obtained from the analyses, where one finds $\Delta E_1 \approx 60\ kJ/mol$, $59\ kJ/mol$, $56\ kJ/mol$, $30\ kJ/mol$, and $24\ kJ/mol$ for 2E1H, 5M2H, 3,7D1O, 1-BL, and 1-PL, respectively, and $\Delta E_2 \approx 73\ kJ/mol$, $70\ kJ/mol$, $62\ kJ/mol$, $34\ kJ/mol$, and $29\ kJ/mol$ for 2E1H, 5M2H, 3,7D1O, 1-BL, and 1-PL, respectively. A larger alkyl group tends to have a higher $\Delta E_1$ and $\Delta E_2$. The enthalpy of H-bonding formation is $\Delta E = \Delta E_2 - \Delta E_1$, which yields $13\ kJ/mol$, $11\ kJ/mol$, $6\ kJ/mol$, $4\ kJ/mol$, and $5\ kJ/mol$ for 2E1H, 5M2H, 3,7D1O, 1-BL, and 1-PL, respectively. These values agree well with the range of the enthalpy of H-bonding formation.[89] Interestingly, the results also suggest MAs with a larger alkyl group tend to have a higher $\Delta E$. This might be due to a larger activation energy for H-bonding dissociation in MAs with a larger alkyl group, as the hydroxyl groups are segregated from the alkyl groups in the supramolecular structures.

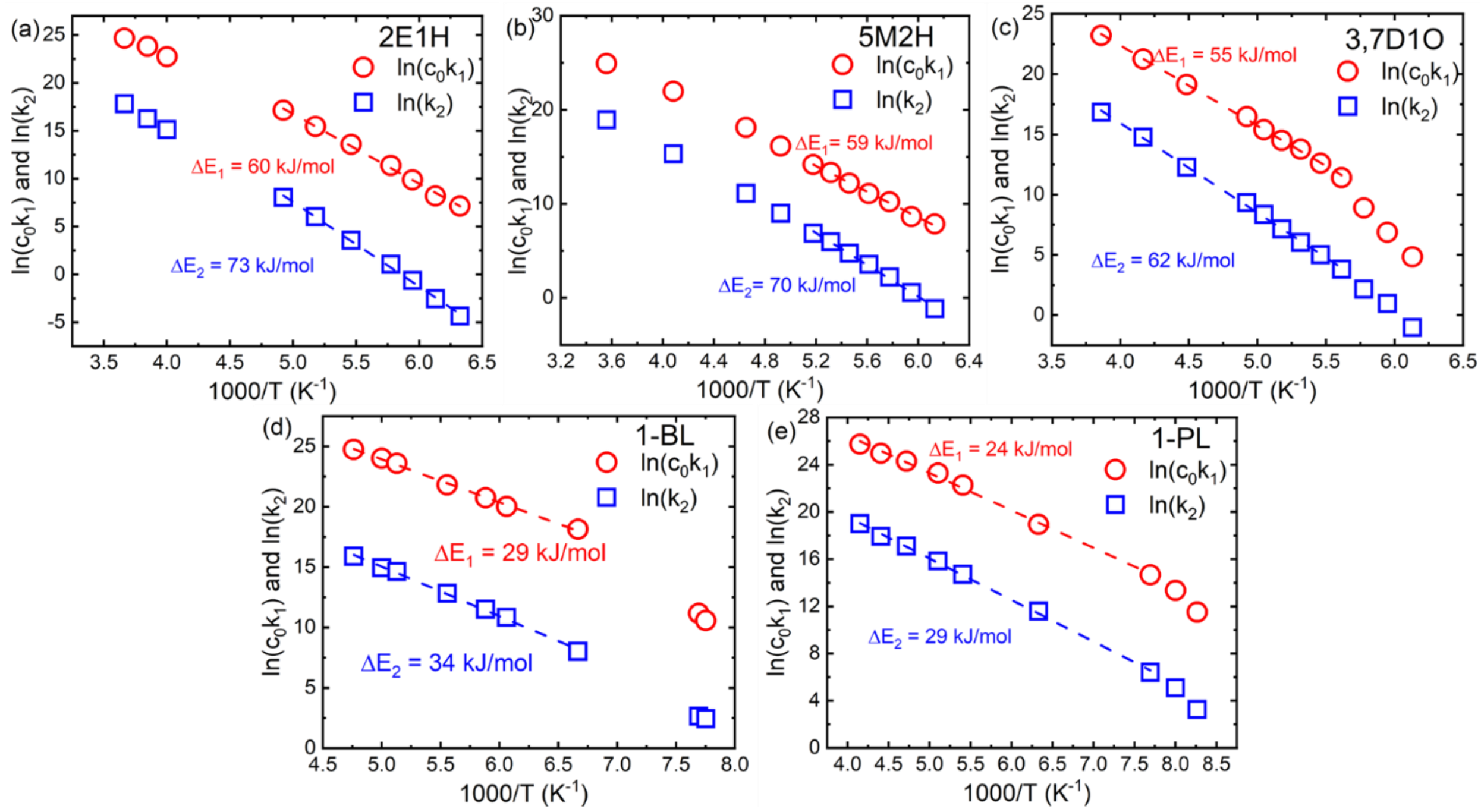

**Figure 14**. Activation plot of $ln\ (c_0k_1)$ and $lnk_2$ of **(a)** 2E1H, **(b)** 5M2H, **(c)** 3,7D1O, **(d)** 1-BL, **(e)** 1-PL. Arrhenius temperature dependences of $ln\ (c_0k_1)$ and $lnk_2$ are observed for all these MAs at high temperatures. The dashed lines represent the Arrhenius fit to the data. The activation energies of H-bonding association and dissociation, $\Delta E_1$ and $\Delta E_2$, are obtained from the Arrhenius fit. Deviations from the Arrhenius temperature dependence are observed at temperatures close to $T_g$ for 3,7D1O, 1-BL, and 1-PL, which is attributed to the strong interplay between glassy dynamics and the reversible H-bonding association/dissociation process.

### 4.5 The supramolecular dynamics of MAs and the Debye relaxation.

The above analyses reveal four characteristic times in the dynamics of MAs with strong supramolecular chains formation: $\tau_H \geq \tau_D \geq \tau_B \geq \tau_\alpha$. In particular, $\tau_H = \tau_\alpha$ defines $\bar{N} = \sqrt{\frac{c_0 K_{eq}(T)}{2}} \approx 1$ or $K_{eq}(T) \approx \frac{2}{c_0}$. As implied from experiments (see **Figure 7** and the schematic **Figure 15**), there are two possible solutions to fulfill $K_{eq}(T) \approx \frac{2}{c_0}$: one at high temperature $T_I$ and the other at low temperature $T_{IV}$. At $T \geq T_I$ or $T \leq T_{IV}$, the H-bonding lifetime is smaller than the structural relaxation. No supramolecular structures form at these temperatures. From this perspective, supramolecular structures exist only at $T_{IV} < T < T_I$. Thus, one can thus define $T_I$ as the highest supramolecular formation temperature (HSFT) above which there is no supramolecular chain formation, and $T_{IV}$ as the lowest supramolecular formation temperature (LSFT) below which the supramolecular structures do not form. We notice that the onset temperature for supramolecular chain formation at high temperatures, $T_I$, has been discussed previously.[44, 45] In contrast, little has been revealed about the onset of supramolecular chain formation in the deep supercooled region at $T_{IV}$. In addition, $\tau_B = \tau_\alpha$ defines $N_R = 1$ that can take place at a high temperature $T_{II}$ and at a low temperature $T_{III}$ (**Figure 15**, also see experimental results of **Figure 7**). Only at $T_{III} < T < T_{II}$, the dynamics sizes, $N_R$, contain more than one monomer unit. As a result, the intermediate dynamics associated with chain breakage and chain swapping, $\tau_B$, appear. Therefore, one can divide the dynamics of MAs into five different dynamics regimes:

**(i) Regime I** at $T \geq T_I$. In this regime, active H-bonding interaction exists. However, $\tau_H$ is shorter than $\tau_\alpha$, $\tau_H \leq \tau_\alpha$. MAs in this region behave like normal polar liquids. The structural relaxation of MAs dictates the terminal flow of MAs. $T_I$ might be higher than or lower than the boiling point, $T_B$, of MAs. If $T_I \geq T_B$, one does not see **Regime I** in the condensed phase and the H-bonding association affects the boiling transition. If $T_I < T_B$, one could observe **Regime I** in the liquid phase and the H-bonding interaction plays a secondary role in controlling the boiling transition.

**(ii) Regime II** at $T_{II} \leq T < T_I$. Supramolecular structures are present, $N_R = 1$, $\tau_B \approx \tau_\alpha$, and $\bar{N} = \tau_D/\tau_\alpha = \tau_H/\tau_\alpha$. In this region, chain swap takes place at a time scale of $\tau_\alpha$. The end-to-end vector reorientation of the supramolecular chain is achieved through active chain swap at the monomer level (**Figure 15**). The Debye relaxation time, $\tau_D$, is proportional to $\bar{N}$, and $\tau_H$ controls the Debye relaxation and the viscosity. It is worth noting that Yuri and Feldman proposed a concert mode of the supramolecular chain structures for the Debye process at high temperatures, which involves a sequential reorientation of supramolecular structures of MA molecules.[32] Yuri et al's picture is very close to the one presented here at $T_{II} \leq T < T_I$, except that the LCM emphasizes active chain swapping. Furthermore, the conclusion that H-bonding lifetime controls the Debye process, which also explains the near Arrhenius temperature dependence of Debye relaxation at this temperature regime (**Figure 7**). Note that the Debye process follows super-Arrhenius temperature dependence at $T < T_{II}$, leading to a kink in the temperature dependence of $\tau_D$ at $T \approx T_{II}$. One can thus estimate $T_{II}$ from the kink of the temperature dependence of $\tau_D$ or viscosity $\eta$ at high temperatures. From this perspective, the activation energy of the Arrhenius temperature dependence of $\tau_D$ or $\eta$ at $T > T_{II}$ provide a direct estimate of the activation energy for H-bond dissociation. In particular, one finds a progressive increment in the apparent activation energy of $\tau_D$ from 20 kJ/mol to 44 kJ/mol at high temperatures from methanol to 1-octanol based on the literature results from Ref. 6 (see **Table S1** of **SM**). In particular, the apparent activation energy of 1-PL and 1-BL is ~30 kJ/mol and ~34 kJ/mol respectively from $\tau_D$ at high temperatures, both of which agree well with the above analyses of ~29 kJ/mol and ~34 kJ/mol from $\tau_H$ of 1-PL and 1-BL, further supporting the above LCM analysis.

**(iii) Regime III** at $T_{III} < T < T_{II}$. In this region, $\tau_H \gg \tau_\alpha$ and $N_R > 1$. Large supramolecular chain structures form. An important feature is the separation in time scales of $\tau_B$ and $\tau_\alpha$. The LCM model has been initially developed to address the dynamics in this regime (**Figure 15**).[16, 17] No chain breakage takes place and the supramolecular chain with length $\bar{N}$ behaves like covalently bonded polymers at $t < \tau_B$. At $t > \tau_B$, chain breakage and chain swap take place. The Debye process is from the end-to-end vector reorientation through a sequential reorientation of dynamics sizes with $N_R$ molecules via chain swapping (**Figure 15**), which takes place at a time scale $\tau_D = \tau_\alpha N_R \bar{N}$. Furthermore, the H-bond lifetime in this region is $\tau_H = \tau_B \bar{N} = \tau_D N_R > \tau_D$, i.e. the H-bond lifetime is longer than the Debye relaxation time in this region. $\tau_H$ follows an Arrhenius temperature dependence with the same activation energy as in **Region II** over a wide temperature window, while $\tau_D$ follows a super-Arrhenius temperature dependence.

Furthermore, deviations of $\tau_H$ from Arrhenius temperature dependence are observed at deep supercooled region around $T_{III}$ (see **Figure 7** and **Figure 15**), signifying a strong influence of molecular cooperativity (configurational entropy) on the H-bonding dissociation and association process. This suggests a crossover from a reaction-controlled process to a diffusion-controlled process of the H-bonding association/dissociation upon cooling. The rapid rise in the configurational entropy barrier of structural relaxation upon cooling leads to a sharp rise in $\tau_\alpha$ and a quick approaching of $\tau_\alpha$ to $\tau_D$ (see **Figure 7** and **Figure 15**). As a result, one should anticipate a reduction in $\bar{N}$ at some intermediate temperature between $T_{II}$ and $T_{III}$, which offers an explanation to the non-monotonic temperature dependence of $\bar{N}$ (**Figure 13**).

**(iv) Regime IV** at $T_{IV} < T \leq T_{III}$. In this region, $\tau_D = \tau_H$, $N_R = 1$, $\tau_B = \tau_\alpha$, and $\bar{N} = \tau_D / \tau_\alpha$. Different from **Regime II**, the influence of molecular diffusion on the H-bonding association and dissociation becomes a dominant factor in this regime. A large configurational entropy barrier exists for the molecular diffusion that leads to large deviations of $\tau_H$ and $K_{eq} = k_1/k_2$ from the Arrhenius temperature dependence. A strong reduction in supramolecular chain length $\bar{N}$ takes place in this regime as well.

**(v) Regime V** at $T \le T_{IV}$. The influence of energy barriers required for molecular cooperativity dominates the H-bonding association and dissociation. No supramolecular chain form and structural relaxation dictate the dynamics of MAs.

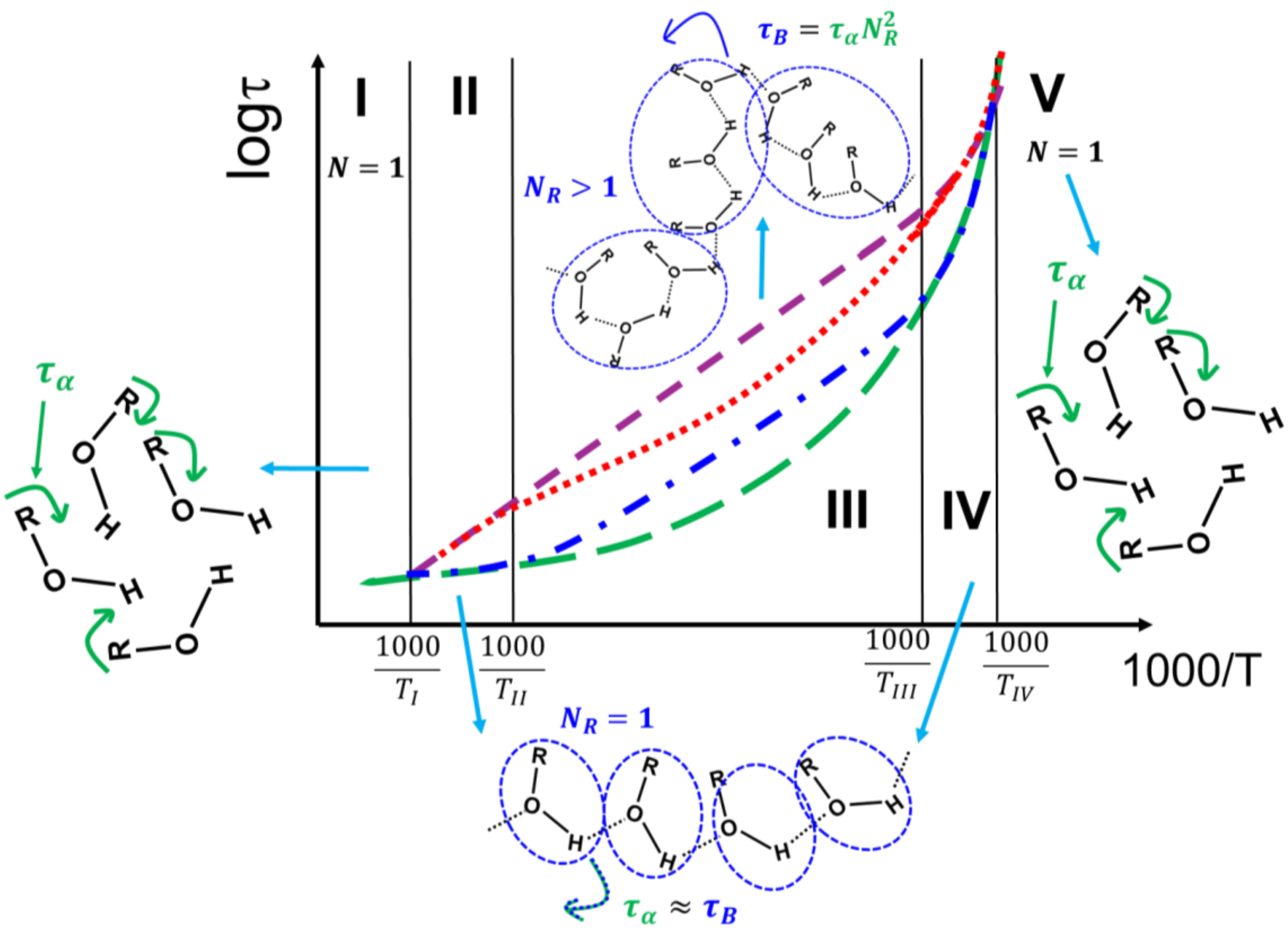

**Figure 15**. A sketch on the relationship between the structural relaxation time $\tau_\alpha$ (green long-dashed line), the chain breakage and swapping time $\tau_B$ (the blue dash-dotted line), the Debye time $\tau_D$ (the red-dotted line), and the hydrogen bonding lifetime $\tau_H$ (the purple short-dashed line), of MAs of supramolecular chain structures. Five dynamics regimes can be distinguished from the relationship between $\tau_\alpha$, $\tau_B$, $\tau_D$, and $\tau_H$. Regime I: $T > T_I$, $\bar{N} = 1$, and no supramolecular chain formation. Regime II: $T_{II} < T \le T_I$, $\bar{N} > 1$, and $N_R = 1$. Supramolecular chain structures form, the dynamics size is $N_R = 1$, and the active chain swapping takes place at the time scale of the structural relaxation, $\tau_\alpha$. Regime III: $T_{III} < T \le T_{II}$, $\bar{N} > 1$, and $N_R > 1$. A non-monotonic temperature dependence of both $\bar{N}$ and $N_R$ are observed upon cooling in this regime. Regime IV: $T_{IV} < T \le T_{III}$, $\bar{N} > 1$, and $N_R = 1$. Regime V: $T < T_{IV}$, $\bar{N} = 1$, and no supramolecular chain formation. Note that the glass transition temperature, $T_g$, might fall in Regimes III, IV, or V. Similarly, the boiling point, $T_B$, might fall into Regimes I or II. When $T_g$ and $T_B$ are between $T_I$ and $T_{IV}$, an influence of the supramolecular structures formation on them should be anticipated. The green arrows represent the monomer or repeat unit reorientation for structural relaxation. The blue circles represent the dynamics sizes, whose reorientation dynamics are represented by the blue arrows that take place at the same time scale of the chain breakage and swapping.

### 4.6 The potential influence of supramolecular structures on the glass transition.

Glass transition might fall within some of the dynamics regions, such as **Regimes III**, **IV**, or **V**. This might introduce additional non-equilibrium effects to the dynamics of MAs. For instance, $T_g$s of 2E1H, 5M2H, and 3,7D1O are higher than their $T_{III}$ (see **Figure 7**), i.e., $T_g > T_{III}$. The low temperature part of **Regime III** and the whole **Regime IV** and **Regime V** of these three MAs are below $T_g$. Similarly, $T_g$s of 1-BL and 1-PL are in **Regime IV** with $T_{IV} < T_g < T_{III}$. MAs form supramolecular structures in both **Regimes III** and **IV**, albeit their different chain sizes. Experimentally, we have not yet observed a MA at $T_g < T_{IV}$, where the structural relaxation of individual MA controls their $T_g$. Since $T_g$ of oligomers depends strongly on their molecular weights, the formation of supramolecular chains should impact strongly the glass forming behaviors of these MAs. In other words, one will need to consider the supramolecular structures of MAs to properly describe their glass formation behaviors and properties at times slower than the structural relaxation, such as viscosity.

Furthermore, if the glass transition intercepts with **Regimes III** or **IV**, rich supramolecular dynamics should appear at temperatures below $T_g$. It would be interesting to characterize the relationship between these supramolecular dynamics and physical aging. Furthermore, a recent study suggest non-hydrogen bonding dipolar liquids can exhibit supramolecular dynamics falling within the framework of LCM.[40] The molecular assemblies have also been actively discussed in model systems such as the Stockmayer fluids and other types of dipolar fluids.[46, 90] Thus, we anticipated the revealed supramolecular dynamics to have strong implications to the understanding of glass transition and dynamics in the deep glassy state of polar liquids.

**4.7 The Kirkwood-Fröhlich factor.**

The Kirkwood-Fröhlich factor, $g_k$, of MAs can be obtained through:

$$g_k = \frac{3(\varepsilon_s - \varepsilon_\infty)k_B T M_0 \varepsilon_0}{F\langle\mu^2\rangle\rho N_A} \qquad (19)$$

where $\varepsilon_s$ is the static dielectric constant and $\mu$ the average dipole moment of the MA. For MA forming supramolecular structures, $\mu = \left(\mu_{||}^2 + \mu_{\perp}^2\right)^{1/2}$ with $\mu_{||}$ and $\mu_{\perp}$ being the characteristic dipolar moment of MA along and perpendicular to the backbone of the suparmolecular structure, respectively. For MAs, we choose $\mu_{||} \approx 1.50\ D$ and $\mu_{\perp} \approx 0.74\ D$ in the analysis.[1]

According to the LCM, $\varepsilon_s - \varepsilon_\infty = \Delta\varepsilon_D + \Delta\varepsilon_L + \Delta\varepsilon_S + \Delta\varepsilon_\alpha$, where $\Delta\varepsilon_D = \epsilon\phi(N > N_R)\,\Delta\varepsilon_{\bar{N}}$, $\Delta\varepsilon_L = \frac{8\epsilon\phi(N>N_R)}{\bar{N}^2\pi^2}\left(\sum_{i=1}^{N_R} i^2\right)\Delta\varepsilon_{\bar{N}}$, $\Delta\varepsilon_S = \Delta\varepsilon_{\bar{N}}\sum_{i=1}^{N_R}\left(\frac{8\epsilon\phi(i)}{\pi^2}\sum_{p=1,odd}^{i}\frac{1}{p^2}\right)$, and $\Delta\varepsilon_\alpha = \frac{\rho N_A\langle\mu_\perp^2\rangle F g_\alpha}{3\varepsilon_0 k_B T M_0} = \frac{\langle\mu_\perp^2\rangle g_\alpha}{\langle\mu_\parallel^2\rangle C_\infty g_{\bar{N}}}\Delta\varepsilon_{\bar{N}}$, where $\Delta\varepsilon_{\bar{N}} = \frac{\rho N_A\langle\mu_\parallel^2\rangle F C_{\bar{N}} g_{\bar{N}}}{3\varepsilon_0 k_B T M_0}$. Thus, one can obtain

$$g_k = \frac{\langle\mu_\parallel^2\rangle C_{\bar{N}} g_{\bar{N}}}{\langle\mu^2\rangle}\left(\epsilon\phi(N > N_R) + \frac{8\epsilon\phi(N > N_R)}{\bar{N}^2\pi^2}\left(\sum_{i=1}^{N_R} i^2\right) + \sum_{i=1}^{N_R}\left(\frac{8\epsilon\phi(i)}{\pi^2}\sum_{p=1,odd}^{i}\frac{1}{p^2}\right) + \frac{\langle\mu_\perp^2\rangle g_\alpha}{\langle\mu_\parallel^2\rangle C_{\bar{N}} g_{\bar{N}}}\right) \quad (20)$$

In particular, $\sum_{i=1}^{N_R}\left(\frac{8\epsilon\phi(i)}{\pi^2}\sum_{p=1,odd}^{i}\frac{1}{p^2}\right) \approx \sum_{i=1}^{N_R}\left(\frac{8\epsilon\phi(i)}{\pi^2}\right) \approx \epsilon\phi(N \le N_R)$. Thus, $\epsilon\phi(N > N_R) + \sum_{i=1}^{N_R}\left(\frac{8\epsilon\phi(i)}{\pi^2}\sum_{p=1}^{i}\frac{1}{p^2}\right) \approx \epsilon\phi(N > N_R) + \epsilon\phi(N \le N_R) = \epsilon$ with $\epsilon$ being the volume fraction of the supramolecular chains. For MAs with strong supramolecular chain formation, **Eqn. 20** can be reduced to

$$g_k \approx \frac{\langle\mu_\parallel^2\rangle\epsilon C_{\bar{N}} g_{\bar{N}} + \langle\mu_\perp^2\rangle g_\alpha}{\langle\mu^2\rangle} \qquad (21)$$

From this perspective, an accurate description of the average $g_k$ of MAs should involve a determination of $\epsilon C_{\bar{N}} g_{\bar{N}}$ and $g_\alpha$.

Experimentally, one observes a strong reduction in $g_k$ with heating. For instance, $g_k$ of 2E1H reduces from $\sim 4.5$ at $T = 160\ K$ to $\sim 3.0$ at $T = 240\ K$. For a given MA and its associated supramolecular structures, both $C_{\bar{N}} g_{\bar{N}}$ and $g_\alpha$ should not exhibit large dependence on temperature. This indicates the strong reduction of $g_k$ with heating is primarily due to a reduction of $\epsilon$, which is consistent the reduction of $\bar{N}$ upon heating. Furthermore, at low temperatures and large $\bar{N}$, $\epsilon \approx 1.0$ holds. **Eqn. 21** can be rewritten as

$$g_k \approx \frac{\langle\mu_\parallel^2\rangle\epsilon C_{\bar{N}} g_{\bar{N}} + \langle\mu_\perp^2\rangle g_\alpha}{\langle\mu^2\rangle} \approx \frac{\langle\mu_\parallel^2\rangle\epsilon C_{\bar{N}} g_{\bar{N}}}{\langle\mu^2\rangle} \qquad (22)$$

Here, we adopt $\langle\mu_\parallel^2\rangle\epsilon C_{\bar{N}} g_{\bar{N}} \gg \langle\mu_\perp^2\rangle g_\alpha$ as reflected by the $\Delta\varepsilon_D + \Delta\varepsilon_S \gg \Delta\varepsilon_\alpha$ in the deep supercooled region of MAs. At temperatures $\epsilon \approx 1.0$ for strong supramolecular chain formation, $g_k \approx \frac{\langle\mu_\parallel^2\rangle C_{\bar{N}} g_{\bar{N}}}{\langle\mu^2\rangle}$ holds. Since $C_{\bar{N}}$ is an intrinsic property of the supramolecular chain and $\frac{\langle\mu_\parallel^2\rangle}{\langle\mu^2\rangle}$ remains constant, this analysis suggests a direct correlation between the alcohol chemical structure, the supramolecular chain formation, and $g_k$. Hence, a saturation of $g_k$ to the order of $C_{\bar{N}} g_{\bar{N}}$ is anticipated at temperatures of $\epsilon \approx 1.0$. As shown in **Figure 12** and literature report,[6] $g_k$ of MAs with strong supramolecular chain formation increases upon cooling and tends to saturate at low temperatures, supporting the above analysis.

## 5. Conclusions

In summary, we have developed a unified theoretical framework that connects the reversible H-bonding interactions in monohydroxy alcohol (MA) with the structure, dynamics, dielectric responses, and linear viscoelastic properties of their supramolecular assemblies. The central physical picture is that hydrogen bonds continuously form and break, producing *living chains*. The free reversibility of H-bonding association and dissociation leads to an exponential chain length distribution, $c(N) \sim \exp\left(-N/\bar{N}\right)$, with an average chain length $\bar{N} \sim \left(c_0 K_{eq}\right)^{1/2}$, which defines a critical condition for supramolecular chain formation at $K_{eq}(T) \sim c_0^{-1}$.

The reversibility of H-bonding association/dissociation also leads to rich dynamics in MAs, including the chain breakage time, $\tau_B$, the Debye time, $\tau_D$, and the H-bonding lifetime, $\tau_H$. In particular, $\tau_B \approx \tau_H/\bar{N}$ holds as a breakage of any one of the H-bonds leads to a breakage of the supramolecular structure. At $\tau_\alpha \leq t \leq \tau_B$, the supramolecular chains behave like covalently bonded polymers and follow Rouse dynamics. The longest Rouse mode that can survive in this region is the one with $N_R \approx (\tau_B/\tau_\alpha)^{1/2}$ monomers. The dynamics at this region are thus composed of two contributions: (i) Rouse dynamics of short chains with $N \leq N_R$ and (ii) the fastest $N_R$ Rouse modes of long chains with $N > N_R$.

These motions are directly observed in both dielectric spectroscopy and linear viscoelastic measurements.

At $t > \tau_B$, reversible chain breakage and recombination take place. The Rouse modes longer than $\tau_B$ of the original chain cannot continue executing as the covalently bonded polymer. As a result, the chain breakage truncates the relaxation time distribution at $\tau_B$. At times of $t \approx \tau_B$, Rouse modes with $N_R$ or less monomers accomplish its end-to-end vector relaxation. Thus, one can view $N_R$ as a dynamic size and renormalize the chain of $\overline{N}$ monomers to $\overline{N}/N_R$ units of dynamic sizes. The long-time dynamics are from swapping of chains with $\overline{N}/N_R$ dynamics sizes. Equivalently, one requires $\overline{N}/N_R$ chain swap to achieve the full chain end-to-end vector reorientation time, i.e. the Debye time, $\tau_D \approx \tau_f \approx N_{eff}\tau_B = \tau_B\overline{N}/N_R = \tau_H/N_R$. Equivalently, the lifetime of the dynamics size controls the end-to-end vector reorientation of the whole chain. At the same time, one can write $\tau_D = \tau_B\overline{N}/N_R = \tau_\alpha\overline{N}N_R < \tau_\alpha\overline{N}^2$, indicating the end-to-end reorientation dynamics of supramolecular chain are proportional to $\overline{N}$ rather than $\overline{N}^2$ like in the unentangled covalently bonded polymers. Therefore, the active chain swapping speeds up the end-to-end reorientation dynamics of the chain. Importantly, short chains of $N < N_R$ do not contribute to dynamics at $t > \tau_B$ and the active chain swapping averages out the effect of chain length at $N > N_R$. Each chain swapping gives energy dissipation that leads to a decay in dynamic modulus. Since $\tau_D \sim \overline{N}$ and $G \sim \overline{N}^{-1}$, one has $G(t) \sim t^{-1}$ at $\tau_B < t < \tau_f$, before a Maxwell mode at $t = \tau_f$ representing the terminal flow of the supramolecular chain. Chain swapping leads to only small net dipole change, and no strong dielectric dispersion was observed at $t > \tau_B$ before the Debye process emerges at $\tau_D$.

Quantitative agreement between the living chain model (LCM) and dielectric spectroscopy and linear viscoelastic measurements allows key structural and kinetic parameters to be extracted from dynamics measurements, including $N_R = \sqrt{\tau_B/\tau_\alpha}$, $\bar{N} = (\tau_D^2/(\tau_B\tau_\alpha))^{1/2}$ , $\tau_H = \tau_D N_R = \tau_D\sqrt{\tau_B/\tau_\alpha}$ , $k_2 = \tau_H^{-1} = \tau_D^{-1}(\tau_B/\tau_\alpha)^{-1/2}$ , and $k_1 = 2\bar{N}^2/(c_0\tau_H) = 2\tau_D^2/(c_0\tau_\alpha\tau_B\tau_H)$. These results reveal a direct connection between H-bonding dynamics and dynamics of the chain-like structures in MAs. Activation energies for H-bonding association and dissociation and their dependence on temperature and alkyl group have been obtained. Arrhenius temperature dependence of $k_2$ and $k_1$ were observed at high temperatures, which exhibit deviations from the original Arrhenius temperature dependence due to an increment in the configurational entropy barrier upon cooling. This highlights a switch from a reaction-controlled process to a diffusion-controlled process of the reversible H-bonding association and dissociation. As a result, non-monotonic temperature dependences of $N_R$ and $\bar{N}$ have been observed. The switch from reaction-controlled to diffusion-controlled reversibility of H-bonding interaction also leads to two critical temperatures, $T_I$ and $T_{IV}$ at $\bar{N} = 1$, and two critical temperatures, $T_{II}$ and $T_{III}$ at $N_R = 1$, signifying five different dynamics regions of the dynamics of MAs due to the influence of H-bonding association/dissociation: $\bar{N} = 1$ at $T \geq T_I$ (Regime I) and $T \leq T_{IV}$ (Regime V), $\bar{N} > 1$ and $N_R = 1$ at $T_{II} \leq T < T_I$ (Regime II) and $T_{III} \leq T < T_{IV}$ (Regime IV), and $\bar{N} > 1$ and $N_R > 1$ at $T_{III} < T < T_{II}$ (Regime III).

Beyond capturing the relaxation times, the LCM also quantitatively describes the dielectric amplitudes and viscoelastic moduli. The dielectric strength of the intermediate process in chain-forming MAs arises primarily from the normal modes of short chains ($N < N_R$), while the dynamic modulus at characteristic frequencies provides independent estimates of $\overline{N}$ and $N_R$. The consistency between these independent measurements supports the validity of the framework. Broadly speaking, this work establishes a quantitative link between reversible intermolecular association and macroscopic dynamics in molecular assemblies. Because hydrogen bonding plays a central role in water and in many chemical and biological systems,[37, 91-94] the present framework provides an understanding of how reversible bonding can control molecular assembly and relaxation. Moreover, recent experiments and theoretical analyses have revealed analogous supramolecular dynamics in non–hydrogen-bonded dipolar liquids where reversible intermolecular associations play an intriguing role.[40, 46, 90] More broadly, we anticipate the revealed features of reversibility from the application of LCM to MA systems can provide key insights on the structures and dynamics of more broader classes of associative liquids, and help investigate the relationship between supramolecular structures formation, the multi-scale dynamics, and their glass transition.

**Acknowledgments**

This work was supported, in part, by the Michigan State University Discretionary Funding Initiative (MSU-DFI). S.C. acknowledges support from National Science Foundation under an Award Number NSF-DMR 2211573.

**Author Declarations**

**Conflict of Interest**

The authors have no conflicts to disclose.

**Data Availability**

The data that support the findings of this study are available within the article and its supplementary materials.

**SUPPLEMENTARY MATERIALS**

## Dynamics of monohydroxy alcohols with chain-like structures: Hydrogen bonding lifetime, chain swapping, and Debye process

Shiwang Cheng*, and Shalin Patil

Department of Chemical Engineering and Materials Science, Michigan State University, East Lansing, MI 48824, USA

Author correspondence should be addressed to Shiwang Cheng at chengsh9@msu.edu

### 1. 2-process fit for 37D1O

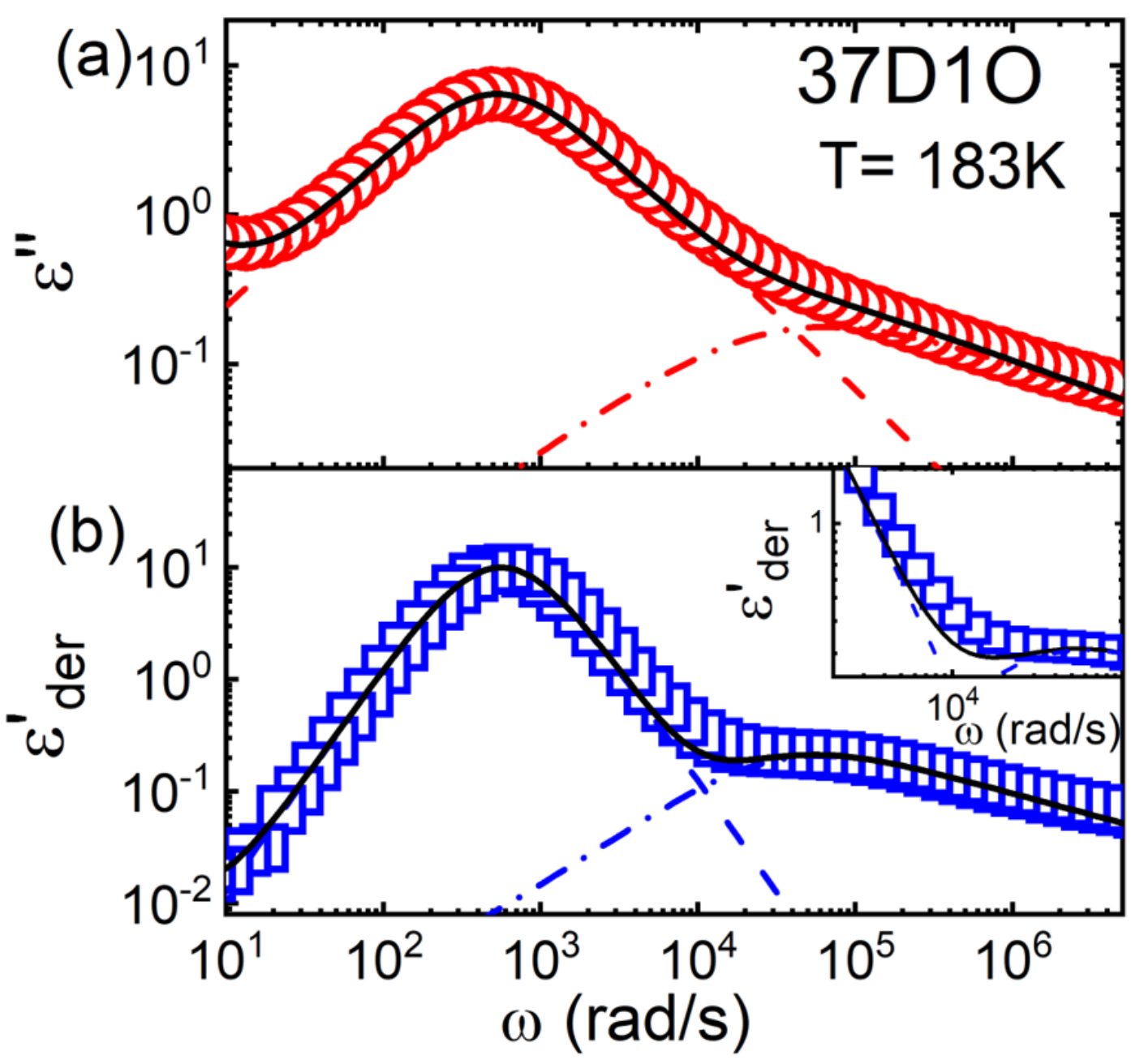

**Figure S1. (a)** The dielectric loss permittivity $\varepsilon''(\omega)$ and **(b)** the derivative spectra $\varepsilon'_{\mathrm{der}}(\omega)$ of 37D1O at 183K. The inset of **(b)** shows the zoomed-in the derivative spectra $\varepsilon'_{\mathrm{der}}(\omega)$ at intermediate frequency.

## 2. Linear Rheology

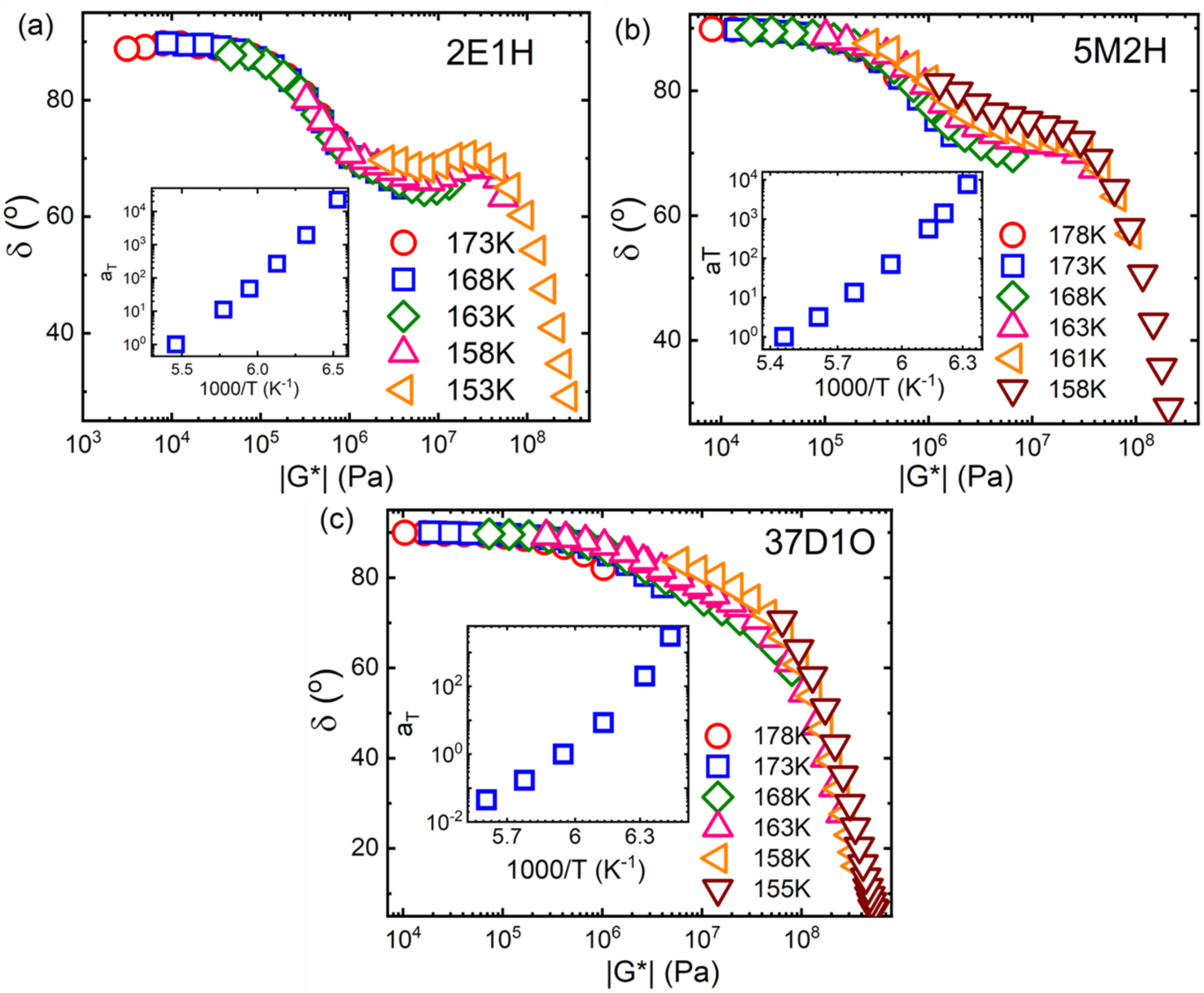

**Figure S2**. (a-c) represents the van Gurp plot for the 2E1H, 5M2H and 3,7D1O respectively. The inset provides the dynamics shift factor, $a_T$, from the linear viscoelastic measurements.

## 3. Comparison of LCM predictions and BDS spectra

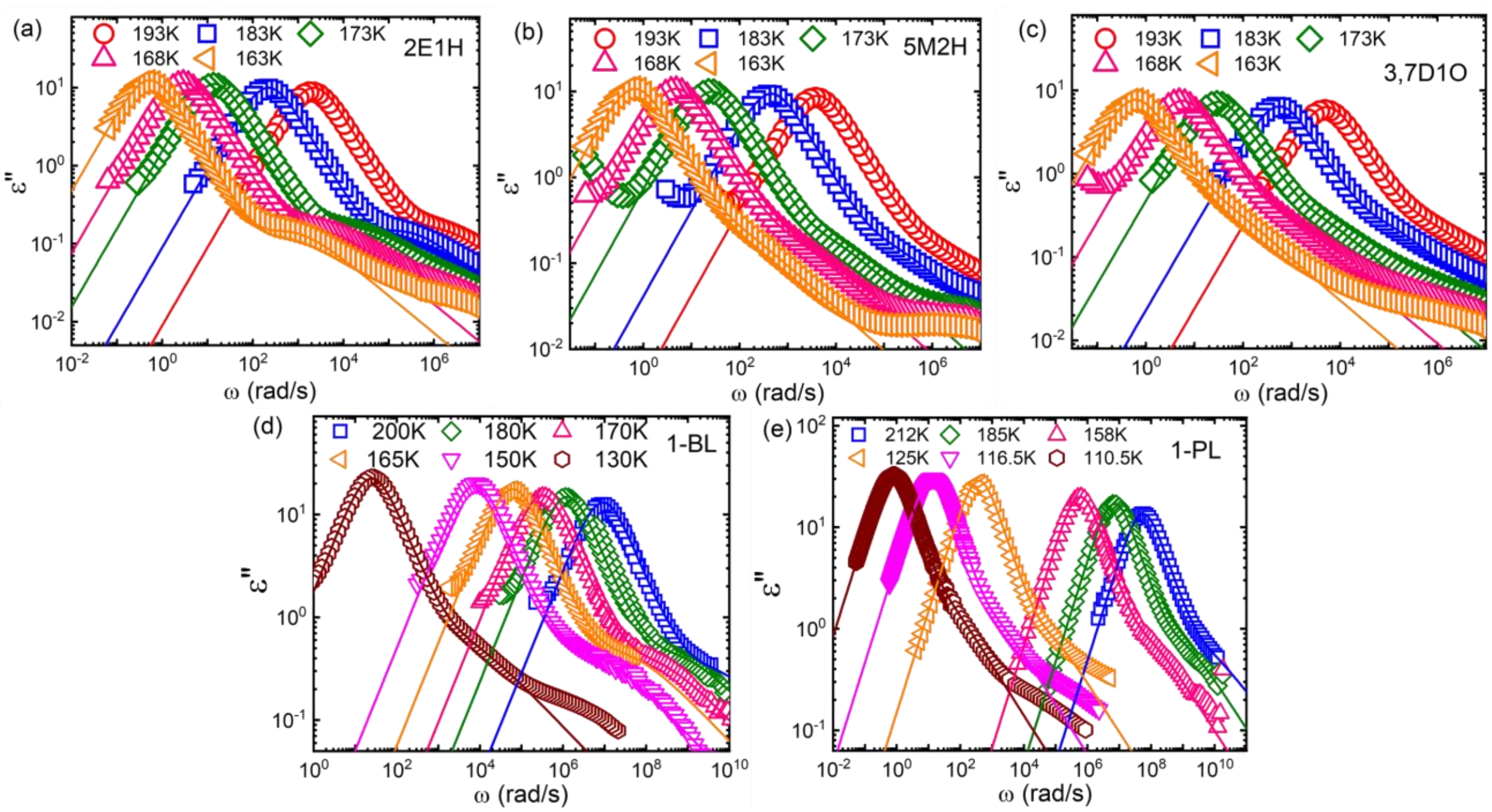

**Figure S3**. Comparison of the predictions of the LCM with the raw dielectric data. The lines represent the contribution of all four processes from LCM predictions.

## 4. Apparent activation energy of $\tau_D$ at high temperatures

**Table S1. Apparent activation energy of $\tau_D$ for different MAs**

| Mono-alcohols | Apparent activation energy for $\tau_D$ (kJ/mol) |
|---|---|
| 1-methanol | $20 \pm 5$ |
| 1-ethanol | $24 \pm 5$ |
| 1-propanol | $30 \pm 5$ |
| 1-butanol | $34 \pm 5$ |
| 1-pentanol | $36 \pm 5$ |
| 1-hexanol | $38 \pm 5$ |
| 1-heptanol | $42 \pm 5$ |
| 1-octanol | $44 \pm 5$ |